\documentclass[graybox]{svmult}

\usepackage{mathptmx}       
\usepackage{helvet}         
\usepackage{courier}        
\usepackage{type1cm}        
\usepackage{float}
\usepackage{url}
\usepackage{makeidx}         
\usepackage{graphicx}        
\usepackage{multicol}        
\usepackage[bottom]{footmisc}

\makeindex             

\begin{document}

\title*{Financial Bubbles, Real Estate bubbles, Derivative Bubbles, and
 the Financial and Economic Crisis}
\author{Didier Sornette and Ryan Woodard}
\institute{Didier Sornette 
\at $^1$ ETH Zurich, Department
of Management, Technology and Economics, Kreuzplatz 5,
CH-8032 Zurich, Switzerland,
\email{dsornette@ethz.ch} \at
$^2$ Swiss Finance Institute, 
c/o University of Geneva, 40 blvd. Du Pont dÕArve
CH 1211 Geneva 4, Switzerland
\and Ryan Woodard \at ETH Zurich, Department
of Management, Technology and Economics, Kreuzplatz 5,
CH-8032 Zurich, Switzerland, \email{rwoodard@ethz.ch}
\at  {\bf JEL classification}:  G01 (Financial Crises), 
G17 (Financial Forecasting),
O16 (Economic Development: Financial Markets; Saving and Capital Investment; Corporate Finance and Governance)
\at {\bf Keywords}: Financial crisis, bubbles, real estate, derivatives, out-of-equilibrium, super-exponential growth,
crashes, complex systems
\at {\bf to appear} in the Proceedings of APFA7 (Applications of Physics in Financial Analysis),
Conference series entitled Applications of Physics in Financial Analysis focuses on the analysis of large-scale Economic data, organized by Misako Takayasu and Tsutomu Watanabe  (\url{http://www.thic-apfa7.com/en/htm/index.html}
}

%
%
\maketitle

\vskip -1cm
\abstract*{The financial crisis of 2008, which started with an initially well-defined epicenter focused on mortgage backed securities (MBS), has been cascading into a global economic recession, whose increasing severity and uncertain duration has led and is continuing to lead to massive losses and damage for billions of people. Heavy central bank interventions and government spending programs have been launched worldwide and especially in the USA and Europe, with the hope to unfreeze credit and boltster consumption.
Here, we present evidence and articulate a general framework that
allows one to diagnose the fundamental cause of the 
unfolding financial and economic crisis: the accumulation of several 
bubbles and their interplay and mutual reinforcement has led to an illusion
of a ``perpetual money machine'' allowing financial institutions to extract
wealth from an unsustainable artificial process. Taking stock of this diagnostic,
we conclude that many of the interventions to address the so-called liquidity crisis
and to encourage more consumption are ill-advised and even
dangerous, given that precautionary reserves were not accumulated in the ``good times''
but that huge liabilities were.
The most ``interesting'' present times constitute unique opportunities
but also great challenges, for which we offer a few recommendations.}

\abstract{The financial crisis of 2008, which started with an initially well-defined epicenter focused on mortgage backed securities (MBS), has been cascading into a global economic recession, whose increasing severity and uncertain duration has led and is continuing to lead to massive losses and damage for billions of people. Heavy central bank interventions and government spending programs have been launched worldwide and especially in the USA and Europe, with the hope to unfreeze credit and boltster consumption.
Here, we present evidence and articulate a general framework that
allows one to diagnose the fundamental cause of the 
unfolding financial and economic crisis: the accumulation of several 
bubbles and their interplay and mutual reinforcement has led to an illusion
of a ``perpetual money machine'' allowing financial institutions to extract
wealth from an unsustainable artificial process. Taking stock of this diagnostic,
we conclude that many of the interventions to address the so-called liquidity crisis
and to encourage more consumption are ill-advised and even
dangerous, given that precautionary reserves were not accumulated in the ``good times''
but that huge liabilities were.
The most ``interesting'' present times constitute unique opportunities
but also great challenges, for which we offer a few recommendations.}


\section{Diagnostics, proximate and systemic origins of the financial crisis}
\label{tktgk3;w}

At the time of writing (first half of April 2009), the World is suffering from a major
financial crisis that has transformed into the worst economic recession since the Great Depression,
perhaps on its way to surpass it. The purpose of the present paper is to relate these developments to
the piling up of five major bubbles: 
\begin{enumerate}
\item  the ``new economy'' ICT bubble starting in the mid-1990s and ending with the crash of 2000, 
\item  the real-estate bubble launched in large part by easy access to a large amount of liquidity as a result of the active monetary policy of the US Federal Reserve lowering the Fed rate from 6.5\% in 2000 to 1\% in 2003 and 2004 in a successful attempt to alleviate the consequence of the 2000 crash,
\item the innovations in financial engineering with the CDOs (collateralized Debt Obligations) and other derivatives of debts and loan instruments issued by banks and eagerly bought by the market, accompanying and fueling the real-estate bubble,
\item the commodity bubble(s) on food, metals and energy, and
\item the stock market bubble peaking in October 2007.
\end{enumerate}
Since mid-2007, the media have been replete with news of large losses by major institutions and by operational and regulatory mishaps. One big question is: how deep will be the losses? Another one is: how severe could be the ensuing recession(s)? 

These questions are stupendous because financial markets have transformed over the past decades from thermometers and liquidity providers of the real economy (tail moving with the dog) into ``the tail wagging the dog,'' 
that is, financial markets now seem to drive the economy.
To mention just one example, there are numerous indications that the corporate strategy of a given firm is significantly influenced by the value of its stock quoted in the capital markets. This is due to many different factors, including incentives (stock options held by CEOs and other top managers), and the financing channels for firm growth offered by higher market valuation, such as during mergers \& acquisition operations (Broekstra et al., 2005).

Our starting point is that financial markets play an essential role in fostering the growth of economies in developed as well as in emergent countries. This impact of financial markets has been growing so much that it is not any longer an exaggeration to suggest that the economy has become in part controlled by a kind of ``beauty contest,'' to paraphrase John Maynard Keynes, where one of the rules of the game for a firm is to appear ``beautiful'' to the financial analysts' eyes and to the investors, by meeting or even beating analysts' earning expectations. In this context, bubbles and crashes exemplify the resulting anomalies.

\subsection{Nature of the financial and economic crisis}

\begin{figure}[H]
\includegraphics[scale=.38]{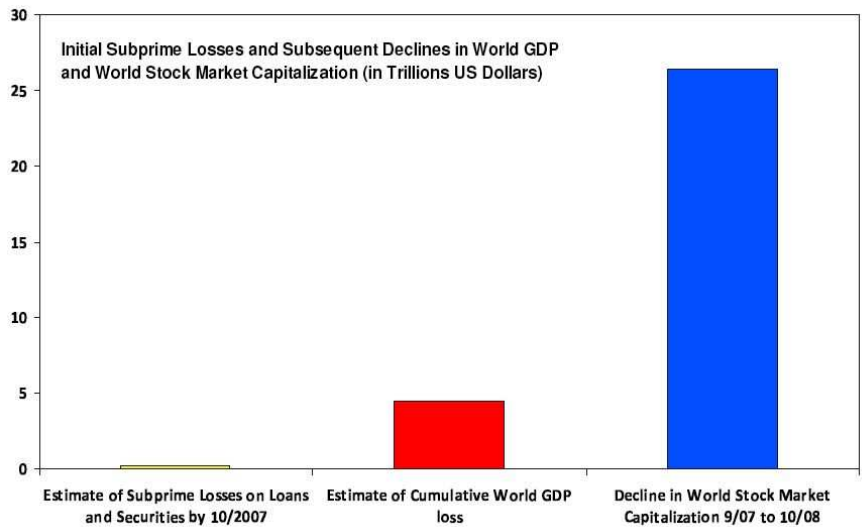}
\caption{Initial subprime losses (almost invisible in the figure) and subsequent declines
up to November 2008 in World GDP
and World stock market capitalization (in Trillions US Dollars).
Source: IMF Global Financial Stability Report; World Economic 
Outlook November update and estimates; World Federation of
Exchanges. Reproduced from Blanchard  (2008)
}
\label{Fig_compare_losses}        
\end{figure}

Better than one thousand words, figure \ref{Fig_compare_losses} compares
the estimated losses for three asset classes \cite{Blanchard08}:
\begin{itemize}
\item losses of U.S. subprime loans and securities, estimated as of October
2007, at about \$250 billion dollars;
\item expected cumulative loss in World output associated with the crisis, based on
forecasts as of November 2008, estimated at \$4,700 billion dollars, that is, about
20 times the initial subprime loss;
\item decrease in the value of stock markets, measured as the sum, over all markets, 
of the decrease in stock market capitalization from July 2007 to
November 2008, estimated at about \$26,400 billion, that is,
100 times the initial subprime loss!  
\end{itemize}

While emphasizing dramatically the cascade from a relatively 
limited and localized event (the subprime
loan crisis in the United States) to the World economy and the
World stock markets, this starting point is deceptive in many ways, 
as will become clear below. The main misconception 
from our viewpoint is reducing the discussion to just the last few years.
The present essay builds an argument that the present turmoil
has its roots going back about 15 years in the past.

\begin{figure}[H]
\includegraphics[scale=.53]{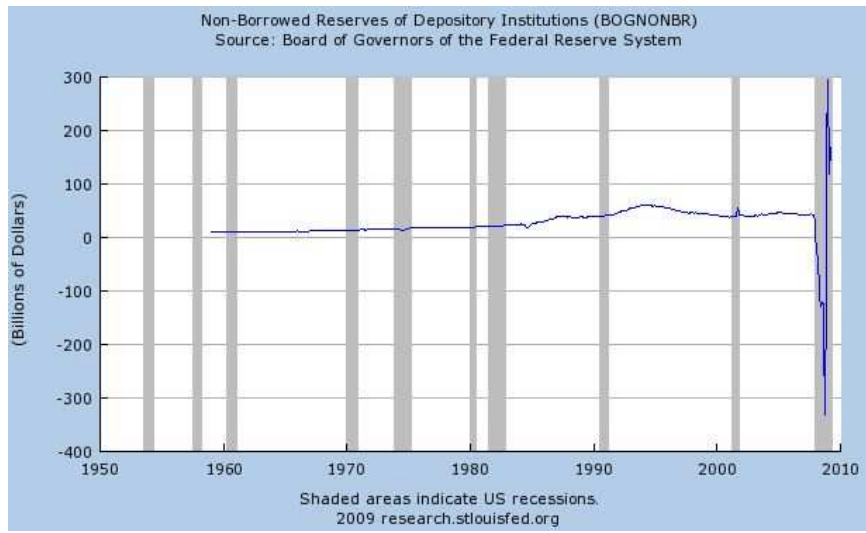}
\caption{Non-borrowed reserves of depository institutions 
from the late 1950s to April 2009. The vertical scale is expressed
in billions of dollars. 
Source: Board of governors of the Federal Reserve system
(2008 Federal Reserve Bank of St. Louis, \protect\url{research.stlouisfed.org})
}
\label{Fig_NonBorrowedReserves}        
\end{figure}

\begin{figure}[H]
\includegraphics[scale=.53]{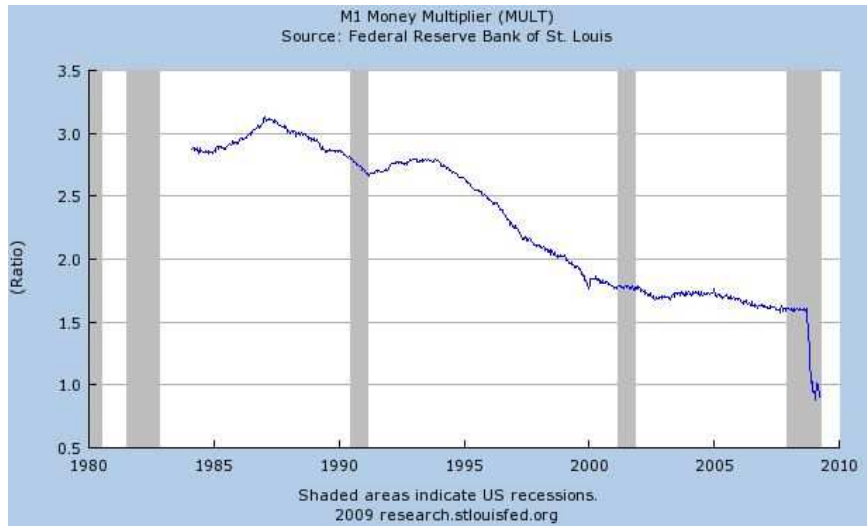}
\caption{Money multiplier M1 defined as
 the ratio of M1 to the St. Louis Adjusted Monetary Base
(\protect\url{http://research.stlouisfed.org/publications/mt/}) 
 from 1983 to March 2009. 
Source: Board of governors of the Federal Reserve system
(Federal Reserve Bank of St. Louis, \protect\url{research.stlouisfed.org})
}
\label{Fig_M1}        
\end{figure}

Figures \ref{Fig_NonBorrowedReserves} and \ref{Fig_M1} provide
additional insights on the extraordinary character of the developments of the present crisis.
First, figure \ref{Fig_NonBorrowedReserves} shows the total amount of
non-borrowed reserves of depository institutions (savings banks
which are regulated by the Federal Deposit Insurance Corporation (FDIC))
from the late 1950s to April 2009. Notice the almost vertical
drop from a level of slightly above $+40$ billion dollars to almost (minus!)  $-350$
billion dollars that occurred in the last quarter of 2008, followed by a
dramatic rebound to $+300$ billion dollars. In the last quarter of 2008, 
under-capitalized banks continued to hemorrhage money via losses
and write-downs of over-valued assets. These banks had to borrow money
from the Federal Reserve to maintain their reserves and their viability.
What is striking in graph \ref{Fig_NonBorrowedReserves}  is
the exceptional amplitudes of the drop and rebound, which 
represent variations completely beyond anything that could have been
foreseen on the basis of the previous 60 years of statistical data. 
Elsewhere, we refer to events such as those shown in figures \ref{Fig_NonBorrowedReserves}
and  \ref{Fig_M1}, which blow up the 
previous statistics, as ``outliers''
\cite{JohansenSornette98, JohansenSornette01} or ``kings'' \cite{LaherSornette}.

Figure \ref{Fig_M1} shows the time evolution of the M1 multiplier, defined as
 the ratio of M1 to the Adjusted Monetary Base estimated by the Federal Reserve
 Bank of St. Louis. Recall that M1 is defined as
the total amount of money\footnote{currency in circulation $+$ checkable deposits (checking deposits, officially called demand deposits, and other deposits that work like checking deposits) $+$ traveler's checks, that is,  
all assets that strictly conform to the definition of money and can be used to pay for a good or service or to repay debt.} in a given country (here the data is for
the U.S.A.). The graph \ref{Fig_M1} again exhibits an extraordinary behavior, with an almost
vertical fall to a level below $1$! This reveals clearly the
complete freezing of lending by financial institutions. Normally, the M1 multiplier is larger than $1$
since money put on a checking account is used at least in part by banks to provide loans. 
The M1 money multiplier has recently slipped below $1$. So each \$1 increase in reserves 
(monetary base) results in the money supply
increasing by \$0.95. This expresses the fact that banks have substantially increased their 
holding of excess reserves while the M1 money supply has not changed by much.
This recent development in the M1 multiplier is another illustration of the extraordinary
occurrence that is presently unfolding.

This concept of ``outliers'' or ``kings'' is important in so far as it stresses the appearance
of transient amplification mechanisms. As we will argue below, the occurrence
of the crisis and its magnitude was predictable and was actually predicted
by some serious independent economists and scholars. They were not taken
seriously at a time when everything seems rosy, leading to what we
refer to as an illusion of the ``perpetual money machine.''
Of course, we am not claiming deterministic predictability for the specific unfolding scenario of the crisis,
only that it was clear that the last 15 years of excesses have led to an unsustainable regime that could
only blow up. In a series of papers to be reviewed below, our group has
repeatedly warned about the succession of bubbles and their unsustainable trajectories
\cite{JohansenSornetteNasdaq00,SorZhou04,zhousornette2004,zhousornette2006,ZhouSor08,SornetteWoodZhou09}.

\subsection{Standard explanations for the financial crisis}

Before we construct our arguments and present the evidence, let us 
review briefly the standard proximal explanations that have been proposed in the
literature. They all share a part of the truth and combine to explain in part
the severity of the crisis. But,  the full extent of the problem can only be
understood from the perspective offered in sections \ref{thrjgvqpfq} and \ref{wthlsjdgb}.

\subsubsection{Falling real estate values}

It has been argued that the immediate cause for the
financial crisis is the bursting of the house price bubble principally in the USA and the
UK and a few other countries, leading to an acceleration of defaults on loans,
translated immediately into a depreciation of the value of mortgage-backed security (MBS)  \cite{Doms07}. 
After a peak in mid-2006 (see subsection \ref{realeatgrgwef}),
the real-estate market in many states plateaued and then started to decrease. 
A number of studies have shown indeed a strong link between house price depreciation and defaults on residential mortgages (see Ref.~\cite{Demyanykreport} and references therein). In particular,  Demyanyk and van Hemert
(2008) \cite{Demyanyk_vanHemert08} explain that all along since 2001
subprime mortgages have been very risky, but their  true riskiness was
hidden by rapid house price appreciation, allowing
mortgage termination by refinancing/prepayment to take place. Only when 
prepayment became very costly (with zero or negative equity in the
house increasing the closing costs of a refinancing), did
defaults took place and the unusually
high default rates of 2006 and 2007 vintage
loans occurred.

The explanation of the crisis based on falling real estate prices is both right and wrong:
right mechanically as understood from the previous paragraph; wrong because
it takes as exogenous the fall in house prices, which would suggest that it comes
as a surprise. In contrast, section \ref{wthlsjdgb} will argue that the fall in real estate
value occurred as part of a larger scheme of events, all linked together.

\subsubsection{Real-estate loans and MBS as a growing asset class held by financial institutions}
\label{hjgrld}

\begin{figure}[H]
\includegraphics[scale=.45]{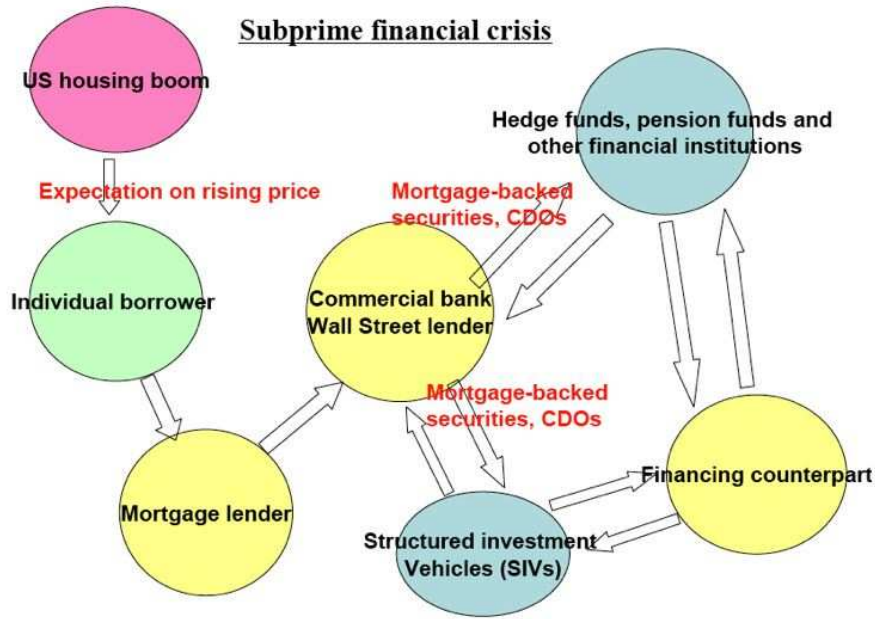}
\caption{Securitization, a form of structured finance, involves the pooling of financial assets, especially those for which there is no ready secondary market, such as mortgages, credit card receivables, student loans. The pooled assets are transfered to a special purpose entity and serve as collateral for new financial assets issued by the entity. The diagram shows the many involved parties.
}
\label{Fig_subprime_network_diagram}        
\end{figure}

A mortgage-backed security (MBS) is a pool of home mortgages that creates 
a stream of payments over time paid
to its owner. The payments are taken from those produced by borrowers who have to
service the interests on their debts.  Figure \ref{Fig_subprime_network_diagram}  
summarizes the network of agents interacting to give life to the MBS.

Developing along with the real-estate bubble, the explosive exponential growth of
the nominal market value of all MBS issued from 2002 to 2007, together with its
subsequent collapse, justifies refering to it as a ``bubble.'' 
According to the Securities Industry and Financial Markets Association, aggregate global CDO 
(collateralized Debt Obligations) issuance grew from USD \$150 billion in 2004, to close to USD \$500 billion in 2006, and to \$2 trillion by the end of 2007. 
 From 0.6 trillion dollars, the cumulative notional
value of CDOs grew to 26 trillion dollars at the end of 2006.
This bubble was fueled firstly by the thirst for larger returns for investors in the USA and in the rest of
the World. It was made possible by a wave of financial innovations leading to the
illusion that the default risks held by lenders, principally banks, could be diversified
away. These  innovations in financial engineering include the CDOs and other derivatives of debts and loan instruments eagerly bought by insurance companies, mutual fund companies, unit trusts, investment trusts, commercial banks, investment banks, pension fund managers, private banking organizations and so on. Since 2007, large losses by major institutions and often related operational and regulatory mishaps have been reported.

The sheer size of the nominal value of MBS held in the books of banks, insurance companies
and many other institutions explains in part the amplitude of the crisis: when
the deflation of the real-estate bubble started, the rate of defaults sky-rocketed and the
holders of MBS started to suffer heavy losses. As a consequence, many financial
institutions have found themselves with insufficient equity and capital, leading to
bankruptcies, fire sale acquisitions or bailouts by governments. 

While compelling, this explanation is incomplete because it does not address the question of 
why did the MBS bubble develop. The underlying mechanisms for bubble formation 
are addressed in section \ref{thrjgvqpfq}. This will help us understand the kind
of inevitability associated with the current crisis.

\subsubsection{Managers' greed and poor Corporate governance problem}

It is clear to all observers that banks have acted incompetently in 
the recent MBS bubble by accepting package risks, by
violating their fiduciary duties to the stockholders, and by letting the
compensation/incentive schemes run out of control.

From executives to salesmen and trading floor operators, incentive mechanisms 
have promoted a generalized climate of moral hazard. Justified by the principles of good corporate governance, executive compensation packages have a perverse dark side of encouraging decision makers to favor strategies that lead to short-term irreversible profits for them at the expense of medium and long-term risks for their firm and their shareholders. Even if the number of CEOs facing forced turnover has increased 3 to 4-fold during the past 20 years while, simultaneously, most contractual severance agreements require the forfeiture of unvested options, lump-sum payments and waiving forfeiture rules often compensate for such losses. There is something amiss when the CEOs of Citibank and of Countrywide walk out of the mess they created for their firms with 9 figure compensation packages. It is often the case that firms finally turn out losing significantly more when the risks unravel than their previous cumulative gains based on these risky positions, while the decision makers responsible for this situation keep their fat bonuses. As long as the risks are borne by the firm and not equally by the decision makers, the ensuing moral hazard will not disappear. It is rational for selfish utility maximizers and 
it will therefore remain a major root of future financial crises.

Herding effects amplify the moral hazard factor just discussed. Indeed, performance is commonly assessed on the basis of comparisons with the average industry performance. Therefore, each manager cannot afford to neglect any high yield investment opportunity that other competitors seem to embrace, even if she believes that, on the long run, it could turn out badly. In addition, herding is often rationalized by the introduction of new concepts, e.g. ``the new economy" and new ``real option" valuation during the Internet bubble. And, herding provides a sense of safety in the numbers: how could everybody be so wrong? Evolutionary psychology and neuro-economics inform us that herding is one of the unavoidable consequences of our strongest cognitive ability, that is, imitation.
In a particularly interesting study using  functional
magnetic resonance imaging on consumption decisions 
performed by teenagers, Berns et al. (2009) have recently shown that the anxiety
generated by the mismatch between oneÕs own preferences and othersÕ
motivates people to switch their choices in the direction of the consensus, suggesting that this is
a major force behind conformity.

Greed, anxiety, moral hazard and psychological traits favoring risk taking in finance 
were prevalent in the past and are bound to remain with us
for the foreseeable future. Therefore, the question whether greed and poor
governance was at the origin of the crisis should be transformed into the question
of timing, that is, why these traits were let loose to foster the development of anomalous excesses
in the last few years.

\subsubsection{Poor lending standards and deteriorating regulations and supervision}

Philippon and Reshef provide an informative view on the question posed by the
title of this section, based on 
detailed information about wages, education and occupations to shed light on the evolution
of the U.S. financial sector from 1906 to 2006 \cite{Philippon09}. They find that 
financial jobs were relatively skill-intensive, complex, and highly paid until the 1930s and after the
1980s, but not in the interim period. They find that the determinants of this evolution are that
financial deregulation and corporate activities linked to IPOs and credit risk increase the demand for
skills in financial jobs, while computers and information technology play a more limited role. 
Philippon and Reshef's analysis
shows that wages in finance were excessively high around 1930 and from the mid 1990s until
2006 \cite{Philippon09}. It is particularly interesting to note that these two periods have
been characterized by considerable excesses in the form of many bubbles and crashes.
The last period is particularly relevant to our arguments
presented in section  \ref{wthlsjdgb} over which a succession of $5$ bubbles developed.

Evidence of deteriorating regulations abounds.
Keys et al. (2008) \cite{Keyslaxscreening} found that (observed) lending standards in the subprime
mortgage market did deteriorate; and the main driving force of the deterioration was the securitization
of those loans. Poser (2009) provides important clues on the failures of the U.S.
 Securities and Exchange Commission. Its most visible fault was its
 inability or reluctance to detect the alleged Madoff Ponzi scheme.
 But Poster \cite{Poser09} points out that the decline in SEC's 
regulatory and enforcement effectiveness began three decades ago.
While in part explained by insufficient resources and inadequate staff training,
the main cause of the SEC decline can probably be attributed to the 
growing prevalence of the ethos of deregulation that pervaded the U.S. government \cite{Poser09}.

This ethos is well exemplified by the failure to pass any legislation on financial derivatives.
Going back to the 1990s, Alan Greenspan, supported successively by then Treasury Secretaries Robert Rubin and Laurence Summers, convinced the U.S. Congress to make the fateful decision not to pass any legislation that would have supervised the development and use of financial derivatives, notwithstanding various attempts by legislators and the call from expert financiers of the caliber of Warren Buffet and Georges Soros who warned years before the present crisis about these ``weapons of financial mass destruction". After being one of the most vocal supporters of the self-regulation efficiency of financial markets, Alan Greenspan is now writing in his memoirs that the villains were the bankers whose self-interest he had once bet upon for self-regulation. 

The story would remain incomplete without distinguishing between
the banking system which is highly regulated and the parallel or shadow
banking system which is much less so \cite{Krugman2008}.
In a speech in June 2008, T.F. Geithner (U.S. Treasury secretary since January 26, 2009) said:
``The structure of the financial system changed fundamentally during the
boom, with dramatic growth in the share of assets outside the
traditional banking system. This non-bank financial system grew to be
very large, particularly in money and funding markets. In early 2007,
asset-backed commercial paper conduits, in structured investment
vehicles, in auction-rate preferred securities, tender option bonds and
variable rate demand notes, had a combined asset size of roughly \$2.2
trillion. Assets financed overnight in triparty repo grew to \$2.5
trillion. Assets held in hedge funds grew to roughly \$1.8 trillion. The
combined balance sheets of the then five major investment banks totaled
\$4 trillion.''  

Given the coexisting two banking systems, the regular
system being explicitly guaranteed with strict capital requirements and the
shadow system being implicitly guaranteed with looser capital requirements,
wealth utility maximizing bankers and investors have been naturally
attracted to the second, which provided new ways to get higher yield \cite{Krugman2008}.
Here, the implicit guarantee is that Bear Stearns, AIG and Merrill Lynch, while
not protected by the FDIC, were protected --as the facts showed--  by the belief
that some firms are too big to fail.

\subsubsection{Did the Fed Cause the Housing Bubble?}

As a logical corollary of the previous subsection, several notable economists 
have blamed the Federal Reserve and the U.S. government  for failing to 
recognize that the shadow banking system, because it was serving the same
role as banks, should have been regulated  \cite{Krugman2008}. 
Stanford economist J.B. Taylor goes further by pointing out 
the errors that the Federal Reserve made in creating and fueling
the crisis \cite{TaylorWallStreet,Taylorbook}, starting with the incredible monetary expansion
of 2002-2003 (described more in section \ref{thbjrgwpbgmvq}), followed by
the excesses of the expansion of government-sponsored Fannie Mae and Freddie Mac who were
encouraged to buy MBS. These errors continued with the misguided diagnostic that the crisis
was a liquidity problem rather than one fundamentally due to counter-party risks.

Actually, A. Greenspan, the former Chairman of the Federal Reserve stated
on October 23, 2008 in a testimony to the U.S. Congress, in reply to questions by Congressman H.A. Waxman:
``I made a mistake in presuming that the self-interests of organizations, specifically banks and others, were such as that they were best capable of protecting their own shareholders and their equity in the firms.'' 
Referring to his free-market ideology, Mr. Greenspan added: ``I have found a flaw. I donÕt know how significant or permanent it is. But I have been very distressed by that fact.''
Mr. Waxman pressed the former Fed chair to clarify his words. ``In other words, you found that your view of the world, your ideology, was not right, it was not working,'' Mr. Waxman said.
``Absolutely, precisely,'' Mr. Greenspan replied. ``You know, that's precisely the reason I was shocked, because I have been going for 40 years or more with very considerable evidence that it was working exceptionally well.''
Greenspan also  said he was ``partially" wrong in  the case of credit default swaps,
complex trading instruments meant to act as insurance against default for bond buyers,
by believing that the market could
handle regulation of derivatives without government intervention.

However, in an article in the Wall Street Journal of March 11, 2009, A. Greenspan 
responded to J.B. Taylor by  defending his policy on two arguments: (1) the Fed controls overnight interest rates, but not ``long-term interest rates and the home-mortgage rates driven by them''; and (2) a global excess of savings was ``the presumptive cause of the world-wide decline in long-term rates.'' Neither argument remains solid 
under scrutiny. First, the post-2002 period was characterized by one-year adjustable-rate mortgages (ARMs), teaser rates that reset in, say, two or three years. Five-year ARMs became ``long-term'' money.
The overnight federal-funds rate that the Fed controls substantially influences the rates on such mortgages. 
Second, Greenspan offers conjecture, not evidence, for his claim of a global savings excess. Taylor has cited evidence from the International Monetary Fund (IMF) to the contrary, however. Global savings and investment as a share of world GDP have been declining since the 1970s, as shown by the data in Taylor's book  \cite{Taylorbook}.

\subsubsection{Bad quantitative risk models in banks (Basel II)}

Since mid-2007, an increasing number of economists,
policy-makers and market operators have blamed the Basel II framework
for banks' capital adequacy to be a major cause for the subprime
financial crisis. 

Basel II is the second of the Basel Accords, which provide recommendations on banking laws and regulations issued by the Basel Committee on Banking Supervision. Basel II  was initially published in June 2004, with 
the purpose of creating an international standard that banking regulators can use 
when creating regulations on how much capital banks need to put aside to guard 
against the types of financial and operational risks banks face. The specific goals
of Basel II are to ensure that capital allocation is more risk sensitive, to separate 
operational risk from credit risk, to quantify both types of risks, and to synchronize
economic and regulatory capital.

First, one should point out that the implementation of Basel II was delayed by
different revisions announced on September 30, 2005 by the four US Federal banking agencies (the Office of the Comptroller of the Currency, the Board of Governors of the Federal Reserve System, the Federal Deposit Insurance Corporation, and the Office of Thrift Supervision) \cite{wikibaselII}.
Second, describing the actual role played by the new prudential
regulation in the crisis and discussing the main arguments raised in the
current debate,  Cannata and Quagliariello (2009) \cite{cannata}
discriminate between more constructive criticisms and
weaker accusations and conclude that there are no sound reasons for 
abandoning the philosophy underlying the Basel II framework.

The dotcom and housing bubbles as well as the development of an inflated financial sphere
were actually apparent to many people. While imperfect, the so-called failure of models
has played a relatively limited role in the unraveling of the crisis. More important is the
desire of economists to think ``things are different this time.''  This is reminiscent of the
``new economy'' mantra of the 1920s preceding the crash of Oct. 1929, the
``new economy'' claim of 1962 during the tronic boom preceding a severe downturn
of the stock market and the ``new economy'' sentiment of the 1990's during the 
ITC bubble. Things change, but some things remain the same, such as greed
and the belief that something fundamentally new is happening that calls for
a downward revision of risk assessment. Herding is further amplified
by the political difficulties in acknowledging independently what data tells us.
B. DeLong, P. Krugman and N. Roubini are among those prominent vocal economists who have
been worried about the development of the economy and the unsustainable succession
of bubbles over the last decade, but they did not have the influence to make a 
significant impact on the US Congress or on Main Street (not to speak of Wall Street).
Unfortunately, few see any pressing need to ask hard questions about the sources of profits
when things are doing well. And even fewer will accept the ``pessimistic'' evidence
that the ``dancing'' is going to stop, when all (superficial) evidence points to the contrary.
Furthermore, one little discussed reason for the present crisis was the lack of
adequate education of top managers on risks in all its dimensions and implications.
How does one expect a CEO without risk culture to act on the face
of the contradictory evidence of, on the one hand, a negative recommendation
of the director of its risk management  department and, on the other hand, great short-term potential
gains in a global exuberant market?
These factors, more than the ``bad'' models, were probably the
problem with the use of quantitative models.

\subsubsection{Rating agency failures}
 
Credit rating agencies have been implicated as principal contributors to the credit crunch and 
 financial crisis. They were supposed to create transparency by rating accurately the riskiness of 
 the financial products generated by banks and financial actors. Their rating should have 
 provided the basis for sound risk-management by mortgage lenders and by creators of structured financial products.
 The problem is that the so-called AAA tranches of MBS have themselves exhibited a rate of default
 many times higher than expected and their traded prices are now just a fraction of their face values.
 
 To provide the rating of a given CDO or MBS, 
 the principal rating agencies -- MoodyÕs, Fitch and Standard \& PoorÕs -- used 
quantitative statistical models based on Monte Carlo simulations to 
 predict the likely probability of default for the mortgages underlying the derivatives.
 One problem is that the default probabilities fed into the calculations 
 were in part based on historical default rates derived from the years 1990-2000, 
 a period when mortgage default rates were low and home prices were rising. 
 In doing so, the models could not factor in correctly the possibility of a general 
 housing bust in which many mortgages are more likely to go into default. 
 The models completely missed the possibility of a global meltdown of the real estate
 markets and the subsequent strong correlation of defaults.
 The complexity of the packaging of the new financial instruments added to the
 problem, since rating agencies had no historical return data for these instruments on which to base their risk assessments. In addition, rating agencies may have felt compelled to deliberately inflate their ratings, either to maximise their consulting fees or because the issuer could be shopping for the highest rating.
 
Recently, Skreta and Veldkamp (2009) \cite{Skreta} showed that all these issues
were amplified by one single factor, the complexity of the new CDO and MBS. The
sheer complexity makes very difficult the calibration of the risks from past data and
from imperfect models that had not yet stood the test of time. In addition, the greater
the complexity, the larger the variability in risk estimations and, thus, of ratings
obtained from different models based on slightly different assumptions. In other words,
greater complexity introduces a large sensitivity to model errors, analogous
to the greater sensitivity to initial conditions in chaotic systems.
If the announced rating is the maximum of all realised ratings, it will be a biased signal of the asset's true quality. The more ratings differ, the stronger are issuers' incentives to selectively disclose (shop for) ratings.
Skreta and Veldkamp think that the incentives for biased reporting of the true risks
have been latent for a long time and only emerged when assets were sufficiently complex that regulation was no longer detailed enough to keep them in check. Note that the abilities of ratings manipulation and shopping to affect asset prices only exist when the buyers of assets are unaware of the games being played by the issuer and rating agency.  This was probably true until 2007, when the crisis exploded.

While these elements are important to understand the financial crisis, they treat
the occurrence of the triggering real estate meltdown as exogenous. In addition, the 
extension of the leveraging on the new MBS and CDO derivatives is not explained.
Overall, we need much more to fully grasp the full underpinning factors of the financial crisis.

\subsubsection{Under-estimating aggregate risks}

As explained above, the wave of financial innovations has led to the illusion that the default risks held by lenders, principally banks, could be diversified away. 
This expectation reflects a widely spread misconception that forgets about
the effects of stronger inter-dependencies associated with tighter firm networks.

Recent multidisciplinary research on self-organizing networks \cite{Sornettebookcrash03,Serrano07,Helbing2008,Borgatti2009} has shown unambiguously that loss of variety, lack of redundancy, removal of compartments,
and stronger ties are all recipes for disaster. This is all the more so because the
medium-sized risks are decreased, giving a false impression of safety based
on the illusion that diversification works. And there is the
emergence of an extremely dangerous collective belief that risks have disappeared.
This led to the so-called ``great moderation'' in the fluctuations of GDP growths of
developed economies and to absurd low risk pricing in financial markets in the last
decade.

Due to globalization
and the intricate networks of bank interdependencies (thousands of banks borrow and
lend to each other every day in a complex ballet) \cite{Freixas,Boss03}, the explosively growing losses on
their MBS books and the realization that other banks were in the same situation have
led to a flight for safety. As a consequence, banks have basically stopped inter-bank
lending for fear of defaults of their financial counterparties. Correlatively, banks have
made more rigid their previously lax lending practices into ridiculously stringent procedures
offered to firms and private customers, basically threatening to freeze the real
economy, which is becoming strangled by cash flow problems.

\subsection{The illusion of the ``perpetual money machine''}
\label{tgt2ogt4gt}

The different elements  described above are only pieces of a greater process
that can be aptly summarized as the illusion of the ``perpetual money machine.'' 
This term refers to the fantasy developed over the 
last 15 years that financial innovations
and the concept that ``this time, it is different'' could provide an accelerated wealth increase.
In the same way that the perpetual motion machine is an impossible dream violating the 
fundamental laws of physics, it is impossible for 
an economy which expands at a real growth rate of 2-3 per cent per year to provide a universal profit of 10-15 per cent per year, as many investors have dreamed of (and obtained on mostly unrealized
market gains in the last decade). 
The overall wealth growth rate
has to equate to the growth rate of the economy. Of course, some sectors can exhibit transient
accelerated growth due to innovations and discoveries. But it is a simple mathematical 
identity that global wealth appreciation has to equal  GDP growth. 

\begin{figure}[H]
\includegraphics[scale=.32]{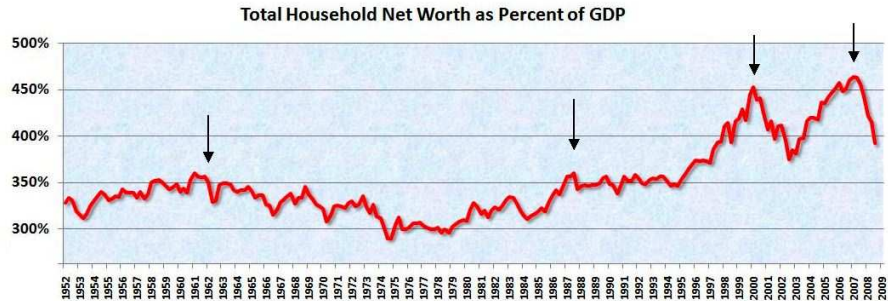}
\caption{Household Net Worth as a percent of GDP from 1952 to March 2009.
This includes real estate and financial assets (stocks, bonds, pension reserves, deposits, etc) net of liabilities (mostly mortgages). The data is from \protect\url{http://www.federalreserve.gov/releases/z1/Current/z1r-5.pdf} (11 Dec. 2008).
Adapted from \protect\url{http://www.calculatedriskblog.com}
}
\label{Fig_NetHousehold_wealth}        
\end{figure}

However, in the
last decade and a half, this identity has been violated by an extraordinary expansion of the
financial sphere. Consider first the evidence given in figure \ref{Fig_NetHousehold_wealth},
which shows the total household net worth in the U.S.  expressed as a fraction of GDP
from 1952 to March 2009. This ratio was relatively stable between 300\% and 350\% for more than 40 years.
Since 1995, two major peaks towering above 450\% can be observed to be followed by their collapse.
The last rightmost arrow points to the peak attained in the third quarter of 2007, which is
followed by a drastic drop. The figure suggests that the drop may have to continue for another 50\%
to 100\% of GDP to come back to historical values. This could occur via a combination
of continuing house value depreciation and stock market losses.

The second peak to the left coincides
with the top of the dotcom bubble in 2000 that was followed by more than two years of strong bearish 
stock markets. The two other arrows to the left, one in 1962 and the other one in 1987
also coincide remarkably with two other bubbles previously documented in the literature:
in 1962, the tronic ``new economy'' bubble collapsed with a cumulative loss of about 35\% in
three months; on 19 October 1987, the famous Black Monday crash occurred that ended
a strong spell of stock market appreciation over the previous few years.

\begin{figure}[H]
\includegraphics[scale=.37]{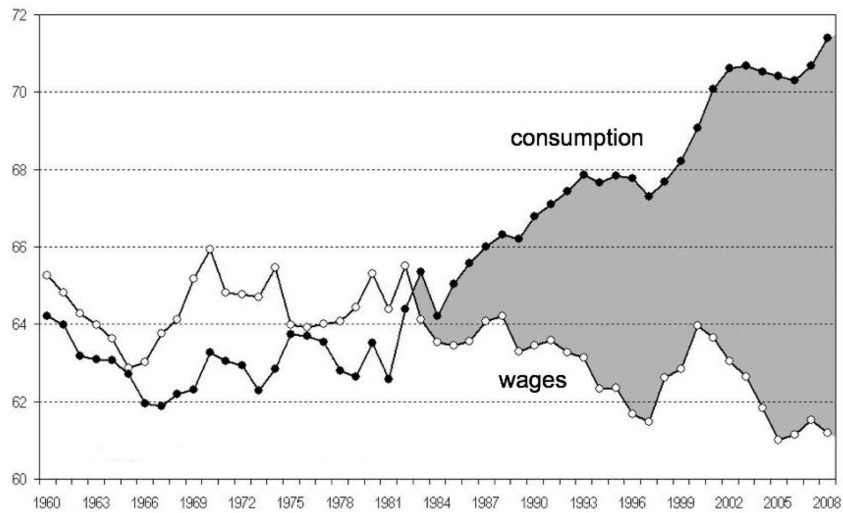}
\caption{Share of wages and of private consumption in Gross Domestic Product (GDP)
for the United States $+$ European Union $+$ Japan.
Source of data and graphics: Michel Husson (http://hussonet.free.fr/toxicap.xls)
}
\label{Fig_consumption-wages}        
\end{figure}

\begin{figure}[H]
\includegraphics[scale=.37]{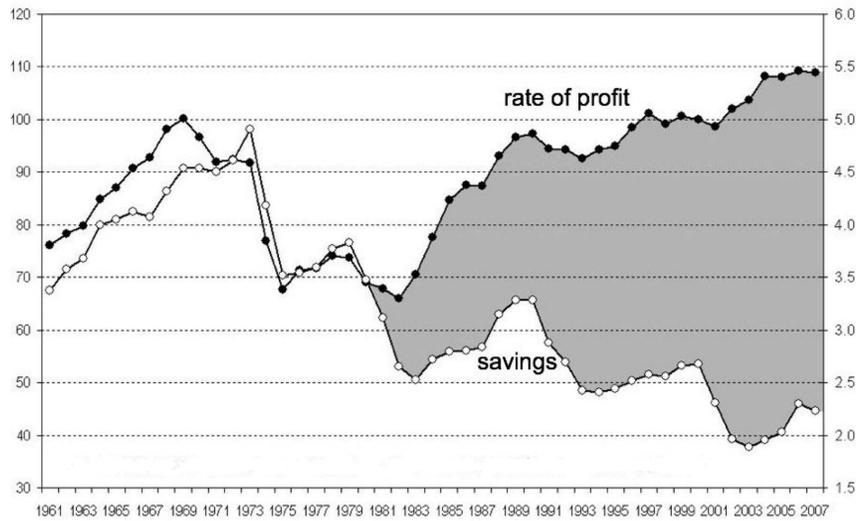}
\caption{Rate of profit (left scale) and rate of accumulation or savings (right scale) for the United States $+$ European Union $+$ Japan.
The rate of accumulation is defined as the rate of growth rate of the net volume of capital $\times$ rate of profit $=$ profit/capital (base: 100 in 2000), Source of data and graphics: Michel Husson (http://hussonet.free.fr/toxicap.xls)
}
\label{Fig_profit_savings}        
\end{figure}

The two figures \ref{Fig_consumption-wages} and \ref{Fig_profit_savings} 
provide another vantage to appreciate fully the impact of the past financial sphere expansion
on the global U.S., European Union and Japan economies. 
First, figure \ref{Fig_consumption-wages}  compares the 
time evolution of private consumption in the U.S.,  European Union and Japan
expressed in percentage of the GDP
to the total wages. One can see that, until 1981, wages funded consumption. 
After 1984, the gap between consumption and wages has been growing 
dramatically. This means of course that consumption had to be funded by
other sources of income than just wages. Figure \ref{Fig_profit_savings} 
suggests that this other source of income is nothing but the increasing profits
from investments, while the diminishing level of savings only partially
covered the increased consumption propensity.
The gap widens between profit and accumulation (gray zones) shown in figure \ref{Fig_profit_savings},
so as to compensate for the difference between
 the share of wages and the share of consumption (gray zones) shown in figure \ref{Fig_consumption-wages}.
 In a nutshell, these two figures
tell us that households in the U.S., 
European Union and Japan have increased their overall level of 
consumption from about 64\% of GDP to almost 72\% of GDP by
extracting wealth from financial profits. Figures for the U.S. alone confirm 
and amplify this conclusion. The big question is whether
the financial profits were translated into real productivity gains
and, therefore, were sustainable. It seems obvious today
to everybody that financial innovations and their profits, which do not provide 
productivity gains in the real
economy, cannot constitute a source of income on the long-term.
This evidence was, however, lost as several exuberant 
bubbles developed during the last 15 years.

\begin{figure}[H]
\includegraphics[scale=.4]{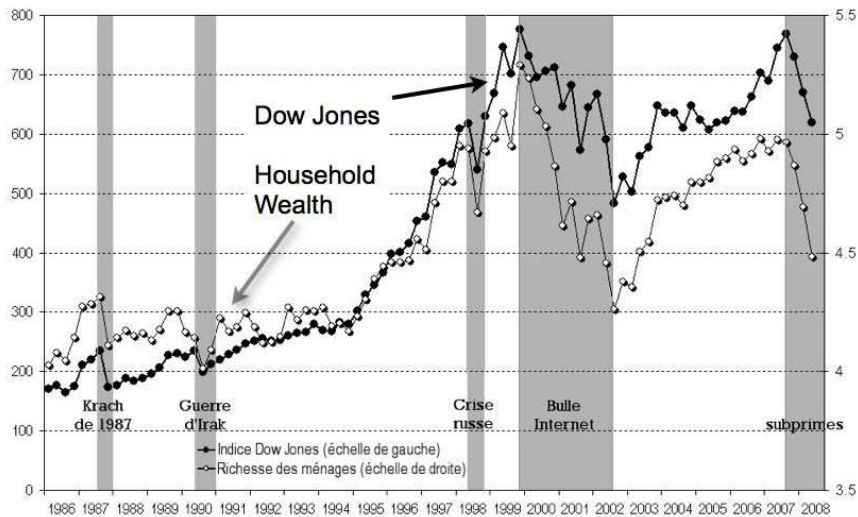}
\caption{The stock market level (left scale) and household wealth in the United States (right scale). 
The Dow Jones Industrial 
Average is shown with base 100 in 1960. The net wealth of households
is given as a multiple of their current income. The five vertical grey zones
outline 5 significant events, which are from left to right: the crash of 1987, 
the Iraq war of 1991, the russian crisis of 1998, the crash and aftermath of
the Internet bubble and the final subprime episode.
Source of data and graphics: Michel Husson (http://hussonet.free.fr/toxicap.xls)}
\label{Fig_household-wealth}        
\end{figure}

The impact of financial profits on the wealth of households is well-illustrated by
figure \ref{Fig_household-wealth}. This graph demonstrates 
the very strong correlation between
U.S. household wealth and the level of the stock market proxied by the Dow Jones
Industrial Average. This supports the concept that financial profits have played
a crucial role in the increase of household consumption discussed above. The component of wealth
due to real estate appreciation during the housing bubble may have actually
played an even bigger role, as it is well documented that the so-called wealth effect
of house value is about twice that of the financial markets \cite{Campbell05}.

As long as the incomes drawn from financial assets are re-invested, the fortunes increase independently of any material link with the real sphere and the variation can potentially increase without serious impediment.
But, financial assets represent the right to a share of the surplus value that is produced. As long as this right is not exercised, it remains virtual. But as soon as anyone exercises it, they discover that it is subject to the law of value, which means one cannot distribute more real wealth than is produced. The 
discrepancy between the exuberant inflation of the financial sphere and the more 
moderate growth of the real economy is the crux of the problem.

The lack of recognition of the fundamental cause of the financial crisis as
stemming from the illusion of the ``perpetual money machine'' is symptomatic of the spirit of the time.
The corollary is that the losses 
are not just the downturn phase of a business or financial cycle. 
They express a simple truth that is too painful to accept for most, that previous gains were not real, 
but just artificially inflated values that have bubbled in the financial sphere, 
without anchor and justification in the real economy. 
In the last decade, banks, insurance companies, Wall Street as well as Main Street and
many of us have lured ourselves into believing that we were richer. 
But this wealth was just the result of a series of self-fulfilling bubbles. As 
explained in more details below, in the USA and in Europe, we had the Internet bubble (1996-2000), the real-estate bubble (2002-2006), the MBS bubble (2002-2007), an equity bubble (2003-3007), and a commodity bubble (2004-2008), each bubble alleviating the pain of the previous bubble or supporting and justifying the next bubble. 

The painful consequence of this brutal truth is that trying to support the level of valuation based on these bubbles is like putting gas in the  ``perpetual money machine.''  Worse, it misuses scarce taxpayer resources, increasing long-term debts and liabilities, which are already at dangerous levels in many countries.

A vivid example is provided by the market valuation of funds
investing in brick-and-mortar companies often observed to be much higher at times of
bubbles than the sum of the value of their components. Objective measures and
indicators can be developed to quantify the ratio of wealth resulting from finance
compared with the total economy. For instance, when it is measured that, on average,
40\% of the income of major US firms result from financial investments, this is clearly
a sign that the US economy is ``building castles in the air'' \cite{Malkiel}.


\section{General framework for bubbles and crashes in finance}
\label{thrjgvqpfq}

\subsection{Introduction}
\label{hntykgte}

Before reviewing the unfolding of the five bubbles over the 15 years that led
to the mother of all crises, we review our approach to the
diagnostic of bubbles and the explanation of crashes. A general review
on models of financial bubbles
encompassing much of the literature can be found in
Ref.~\cite{Kaizoji-Sornette09}.

\begin{figure}[H]
\includegraphics[scale=.4]{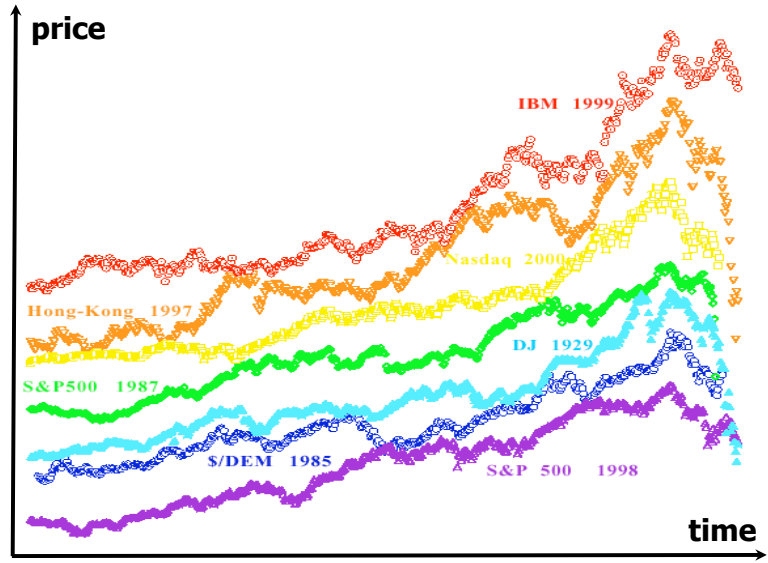}
\caption{Seven bubbles that ended in severe crashes.
The bubble examples include stock market indices, individual companies,
currencies, and for different epochs in the twentieth century.
Each bubble has been rescaled vertically and translated
to end at the time of the crash on the right of the graph. The horizontal axis covers approximately
2.5 years of data. The legend for each of the seven bubbles indicates the name
of the asset supporting the bubble and the year when the crash occurred.
}
\label{Fig_Bubbles_Butterflies}        
\end{figure}

Consider the seven price trajectories shown in figure \ref{Fig_Bubbles_Butterflies}.
They are seven bubbles that ended in very severe crashes. This figure illustrates
the common future that crashes occur after a spell of very strong value appreciation,
following a similar pattern. This suggests a common underlying mechanism.

According to the consecrated academic view that markets are efficient, only the revelation of a dramatic piece of information can cause a crash, yet in reality even the most thorough post-mortem analyses are typically inconclusive as to what this piece of information might have been. This is certainly true for the seven cases
shown in  figure \ref{Fig_Bubbles_Butterflies} (see Ref.~\cite{Sornettebookcrash03} for a detailed discussion).

Most approaches to explaining crashes search for possible mechanisms or effects that operate at very short time scales (hours, days, or weeks at most). Here, we build on the radically different hypothesis \cite{Sornettebookcrash03} that the underlying cause of the crash should be found in the preceding months and years, in the progressively increasing build-up of market cooperativity, or effective interactions between investors, often translated into accelerating ascent of the market price (the bubble). According to this ``critical'' point of view, the specific manner by which prices collapsed is not the most important problem: a crash occurs because the market has entered an unstable phase and any small disturbance or process may reveal the existence of the instability. Think of a ruler held up vertically on your finger: this very unstable position will lead eventually to its collapse, as a result of a small (or an absence of adequate) motion of your hand or due to any tiny whiff of air. The collapse is fundamentally due to the unstable position; the instantaneous cause of the collapse is secondary. In the same vein, the growth of the sensitivity and the growing instability of the market close to such a critical point might explain why attempts to unravel the proximal origin of the crash have been so diverse. Essentially, anything would work once the system is ripe. 

What is the origin of the maturing instability? A follow-up hypothesis underlying this paper is that, in some regimes, there are significant behavioral effects underlying price formation leading to the concept of ``bubble risks.'' This idea is probably best exemplified in the context of financial bubbles, such as the recent Internet example culminating in 2000 or the real-estate bubble in the USA culminating in 2006. Many studies have suggested that bubbles result from the over-optimistic expectation of future earnings (see, for instance, Ref.~\cite{Sheffrin05}), and many works have argued contrarily for rational explanations (for example, Ref.~\cite{Garber00}). History provides a significant number of examples of bubbles driven by unrealistic expectations of future earnings followed by crashes. The same basic ingredients have been documented to occur repeatedly \cite{Sornettebookcrash03}. According to this view, fuelled by initially well-founded economic fundamentals, investors develop a self-fulfilling enthusiasm by an imitative process or crowd behavior that leads to the building of castles in the air, to paraphrase Malkiel  \cite{Malkiel}. Our previous research suggests that the ideal economic view, that stock markets are both efficient and unpredictable, may be not fully correct. We propose that, to understand stock markets, one needs to consider the impact of positive feedbacks via possible technical as well as behavioral mechanisms, such as imitation and herding, leading to self-organized cooperativity and the development of possible endogenous instabilities. We thus propose to explore the consequences of the concept that most of the crashes have fundamentally an endogenous, or internal, origin and that exogenous, or external, shocks only serve as triggering factors. As a consequence, the origin of crashes is probably much more subtle than often thought, as it is constructed progressively by the market as a whole, as a self-organizing process. In this sense, the true cause of a crash could be termed a systemic instability.

By studying many empirical historical examples, C. Kindleberger has identified the universal scenario associated with the development of bubbles \cite{Kindleberger78} as follows (see also Ref.~\cite{Sornettebookcrash03}):
\begin{equation}
{\rm displacement}  ~\to ~{\rm credit ~creation}~ \to~{\rm euphoria}~ \to~{\rm critical ~financial ~distress}
~\to~ \rm{revulsion}
\label{thy;e;ss}
\end{equation}
The upswing usually starts with an opportunity (``displacement'') --new markets, new technologies or some dramatic political change--  and investors looking for good returns. The scenario proceeds through the euphoria of rising prices, particularly of assets, while an expansion of credit inflates the bubble.
In the manic euphoric phase, investors scramble to get out of money and into illiquid things such as stocks, commodities, real estate or tulip bulbs: a larger and larger group of people seeks to become rich without a real understanding of the processes involved.
Ultimately, the markets stop rising and people who have borrowed heavily find themselves overstretched. This is distress, which generates unexpected failures, followed by revulsion or discredit.
The final phase is a self-feeding panic, where the bubble bursts. People of wealth and credit scramble to unload whatever they have bought at greater and greater losses, and cash becomes king.
The sudden fall, first in the price of the primary object of
speculation, then in most or all assets, is associated with a reverse rush for liquidity. Bankruptcies increase.
Liquidation speeds up, sometimes degenerating into panic. The value of collateral (credit and money) sharply contracts. Then, debt deflation ends as productive assets move from financially weak owners (often speculators or the original entrepreneurs) to financially strong owners (well capitalized financiers). This provides the foundation for another cycle, assuming that all the required factors (displacement, monetary expansion, appetite for speculation) are present.

\subsection{Conceptual framework}

Let us now focus on the empirical question of the existence and detection of financial bubbles. But what are really bubbles? The term ``bubble'' is widely used but rarely clearly defined. Following (Case and Shiller, 2003) \cite{CaseShiller}, the term ``bubble'' refers to a situation in which excessive public expectations of future price increases cause prices to be temporarily elevated. For instance, during a housing price bubble, homebuyers think that a home that they would normally consider too expensive for them is now an acceptable purchase because they will be compensated by significant further price increases. They will not need to save as much as they otherwise might, because they expect the increased value of their home to do the saving for them. First-time homebuyers may also worry during a housing bubble that if they do not buy now, they will not be able to afford a home later. Furthermore, the expectation of large price increases may have a strong impact on demand if people think that home prices are very unlikely to fall, and certainly not likely to fall for long, so that there is little perceived risk associated with an investment in a home.

What is the origin of bubbles? In a nutshell, speculative bubbles are caused by ``precipitating factors'' that change public opinion about markets or that have an immediate impact on demand, and by ``amplification mechanisms'' that take the form of price-to-price feedback, as stressed by Shiller (2000) \cite{Shiller00}. Consider again the example of a housing bubble. A number of fundamental factors can influence price movements in housing markets. On the demand side, demographics, income growth, employment growth, changes in financing mechanisms or interest rates, as well as changes in location characteristics such as accessibility, schools, or crime, to name a few, have been shown to have effects. On the supply side, attention has been paid to construction costs, the age of the housing stock, and the industrial organization of the housing market. The elasticity of supply has been shown to be a critical factor in the cyclical behavior of home prices. The cyclical process that we observed in the 1980s in those cities experiencing boom-and-bust cycles was caused by the general economic expansion, best proxied by employment gains, which drove demand up. In the short run, those increases in demand encountered an inelastic supply of housing and developable land, inventories of for-sale properties shrank, and vacancy declined. As a consequence, prices accelerated. This provided an amplification mechanism as it led buyers to anticipate further gains, and the bubble was born. Once prices overshoot or supply catches up, inventories begin to rise, time on the market increases, vacancy rises, and price increases slow down, eventually encountering downward stickiness. The predominant story about home prices is always the prices themselves \cite{Shiller00,Sornettebookcrash03}; the feedback from initial price increases to further price increases is a mechanism that amplifies the effects of the precipitating factors. If prices are going up rapidly, there is much word-of-mouth communication, a hallmark of a bubble. The word-of-mouth can spread optimistic stories and thus help cause an overreaction to other stories, such as ones about employment. The amplification can also work on the downside as well. 

Another vivid example is the proposition offered close to the peak of the Internet bubble that culminated in 2000, that better business models, the network effect, first-to-scale advantages, and real options effect could account rationally for the high prices of dot-com and other New Economy companies \cite{Mauboussin}. These interesting views expounded in early 1999 were in synchrony with the bull market of 1999 and preceding years. They participated in the general optimistic view and added to the strength of the herd. Later, after the collapse of the bubble, these explanations seemed less attractive. This did not escape U.S. Federal Reserve chairman Alan Greenspan (1997), who said \cite{Greenspan97}: ``Is it possible that there is something fundamentally new about this current period that would warrant such complacency? Yes, it is possible. Markets may have become more efficient, competition is more global, and information technology has doubtless enhanced the stability of business operations. But, regrettably, history is strewn with visions of such new eras that, in the end, have proven to be a mirage. In short, history counsels caution.'' In this vein, as mentioned above, the buzzword ``new economy'' so much used in the late 1990s was also hot in the 1960s during the ``tronic boom'' before a market crash, and during the bubble of the late 1920Õs before the Oct. 1929 crash. In this latter case, the ``new'' economy was referring to firms in the utility sector. It is remarkable how traders do not learn the lessons of their predecessors!

Positive feedback occurs when an action leads to consequences which themselves reinforce the action and so on, leading to virtuous or vicious circles. We propose the hypotheses that (1) bubbles may be the result of positive feedbacks and (2) the dynamical signature of bubbles derives from the interplay between fundamental value investment and more technical analysis. The former can be embodied in nonlinear extensions of the standard financial Black-Scholes model of log-price variations 
\cite{SornetteAndersen02,IdeSornette02,Corcos02,AndersenSornette04}.

The mechanisms for positive feedbacks in financial markets include
(1) technical and rational mechanisms (option hedging, insurance portfolio strategies,
trend following investment strategies, asymmetric information on hedging strategies)
and (2) behavioral mechanisms (breakdown of ``psychological Galilean invariance'' \cite{Sornette08}, 
imitation). We stress here particularly the second mechanism which, we believe, dominates.
First,  it is actually ``rational'' to imitate when lacking sufficient time, energy and information to make a decision based only on private information and processing, that is, most of the time.
Second, imitation has been documented in psychology and in neuro-sciences as one of the most evolved cognitive processes, requiring a developed cortex and sophisticated processing abilities. 
It seems that imitation has evolved as an evolutionary advantageous trait, and may even have promoted the development of our anomalously large brain (compared with other mammals) \cite{Dunbar1998}.
Furthermore, we learn our basics and how to adapt mostly by imitation all through our life.
Imitation is now understood
as providing an efficient mechanism of social learning. 
Experiments in developmental psychology suggest that infants use imitation to get to know people, possibly applying a like-me test (people who I can imitate and who imitate me). Imitation is found in highly social living species which show, from a human observer point of view, intelligent behavior and signs for the evolution of traditions and culture (humans and chimpanzees, whales and dolphins, parrots). 
In non-natural agents as robots, imitation is a principal tool for easing the programming of complex tasks or endowing groups of robots with the ability to share skills without the intervention of a programmer. Imitation plays an important role in the more general context of interaction and collaboration between software agents and human users. 

Humans are perhaps the most social mammals and they shape their environment to their personal and social needs. This statement is based on a growing body of research at the frontier between new disciplines called neuro-economics, evolutionary psychology, cognitive science, and behavioral finance
\cite{Damasio94,Camerer03,Gintisetal05}. This body of evidence emphasizes the very human nature of humans with its biases and limitations, opposed to the previously prevailing view of rational economic agents optimizing their decisions based on unlimited access to information and to computation resources.

Imitation, in obvious or subtle forms, is a pervasive activity of humans. In the modern business, economic and financial worlds, the tendency for humans to imitate leads in its strongest form to herding and to crowd effects. Imitation is a prevalent form in marketing with the development of fashion and brands.
We hypothesize that financial bubbles are footprints of perhaps the most robust trait of humans and the most visible imprint in our social affairs: imitation and herding (see Ref.~\cite{Sornettebookcrash03}, and references therein).

\subsection{Finite-time singular behavior of bubbles}
\label{LPPLtreatg}

\begin{figure}[H]
\includegraphics[scale=.4]{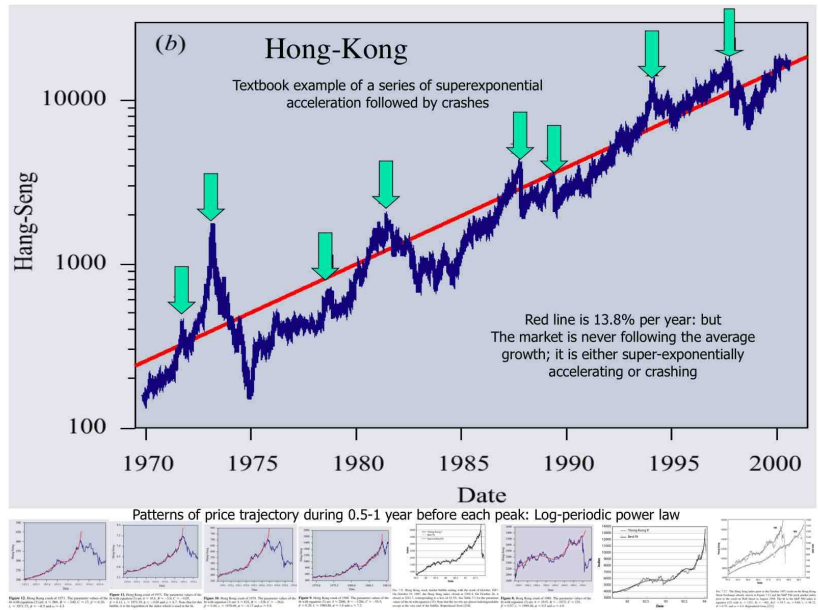}
\caption{Trajectory of the Hong-Kong Hang Seng index from 1970 to 2000. The vertical log-scale together with the linear time scale allows one to qualify an exponential growth with constant growth rate as a straight line. This is indeed the long-term behavior of this market, as shown by the best linear fit represented by the solid straight line, corresponding to an average constant growth rate of 13.8\% per year. The 8 arrows point to 8 local maxima that were followed by a drop of the index of more than 15\% in less than three weeks (a possible definition of a crash). The 8 small panels at the bottom show the upward curvature of the log-price trajectory preceding each of these local maxima, which diagnose an unsustainable bubble regime, which culminates at the peak before crashing.
Reproduced from Ref.~\protect\cite{SorJoh01}
}
\label{Fig_HengSeng_Bubbletextbook}        
\end{figure}

This understanding of imitation and herding
has led us to propose that one of the hallmarks of a financial bubble is the faster-than-exponential growth of the price of the asset under consideration. It is convenient to model this accelerated growth by a power law with a so-called finite-time singularity \cite{Sornetteetal03}. This feature is nicely illustrated by the price trajectory of the Hong-Kong Hang Seng index from 1970 to 2000, as shown in figure \ref{Fig_HengSeng_Bubbletextbook}. The Hong Kong financial market is repeatedly rated as providing one of the most pro-economic, pro-entrepreneurship and free market-friendly environment in the world, and thus provides a textbook example of the behavior of weakly regulated liquid and striving financial markets. In figure \ref{Fig_HengSeng_Bubbletextbook}, the logarithm of the price $p(t)$ is plotted as a function of the time (in linear scale), so that an upward trending straight line qualifies as exponential growth with a constant growth rate equal to the slope of the line: the straight solid line corresponds indeed to an approximately constant compounded growth rate of the Hang Seng index equal to $13.8\%$ per year.  However, the most striking feature of figure \ref{Fig_HengSeng_Bubbletextbook} is not this average behavior, but the obvious fact that the real market is never following and abiding to a constant growth rate. One can observe a succession of price run-ups characterized by growth rates ... growing themselves: this is reflected visually in figure \ref{Fig_HengSeng_Bubbletextbook} by transient regimes characterized by strong upward curvature of the price trajectory. Such an upward curvature in a linear-log plot is a first visual diagnostic of a faster than exponential growth (which of course needs to be confirmed by rigorous statistical testing). Such a price trajectory can be approximated by a characteristic transient finite-time singular power law of the form 
\begin{equation}
\ln[p(t)] = A + B (t_c-t)^m~,~~~~~{\rm where}~ B<0,    ~~0<m<1  ~ ,
 \label{tgtga}
\end{equation}
and $t_c$ is the theoretical critical time corresponding to the end of the transient run-up (end of the bubble). Such transient faster-than-exponential growth of $p(t)$ is our working definition of a bubble. It has the major advantage of avoiding the conundrum of distinguishing between exponentially growing fundamental price and exponentially growing bubble price, which is a problem permeating most of the previous statistical tests developed to identify bubbles (see Ref.~\cite{Lux-Sornette02} and references therein). The conditions $B<0$ and $0<m<1$ ensure the super-exponential acceleration of the price, together with the condition that the price remains finite even at $t_c$. Stronger singularities can appear for $m<0$ \cite{GluzmanSornette02}.

Such a mathematical expression (\ref{tgtga}) is obtained from models that capture the effect of a positive feedback mechanism. Let us illustrate it with the simplest example. Starting with a standard proportional growth process $dp/dt = r p$ (omitting for the sake of pedagogy the stochastic component), where $r$ is the growth rate, let us assume that $r$ is itself an increasing function of the price $p$, as a result of the positive feedback of the price on the future returns. For illustration, let us assume that $r$ is simply proportional to $p$ ($r=c p$, where $c$ is a constant), so that the proportional growth equation become $dp/dt = c p^2$. The solution of this equation is of the form (\ref{tgtga}) where $\ln[p(t)]$ is replaced by $p(t)$,  with $m=-1$ and $A=0$, corresponding to a divergence of $p(t)$ at $t_c$. Many systems exhibit similar transient super-exponential growth regimes, which are described mathematically by power law growth with an ultimate finite-time singular behavior: planet formation in solar systems by runaway accretion of planetesimals, Euler equation of inviscid fluids, general relativity coupled to a mass field leading to formation of black holes in finite time, Zakharov equation of beam-driven Langmuir turbulence in plasma, rupture and material failures, nucleation of earthquakes modeled with the slip-and-velocity weakening Ruina-Dieterich friction law, models of micro-organisms interacting through chemotaxis aggregating to form fruiting bodies, Mullins-Sekerka surface instability, jets from a singular surface, fluid drop snap-off, the Euler rotating disk, and so on. Such mathematical equations can actually provide an accurate description of the transient dynamics, not too close to the mathematical singularity where new mechanisms come into play. The singularity at $t_c$ mainly signals a change of regime. In the present context, $t_c$ is the end of the bubble and the beginning of a new market phase, possible a crash or a different regime.

Such an approach may be thought at first sight to be inadequate or too naive to capture the intrinsic stochastic nature of financial prices, whose null hypothesis is the geometric random walk model \cite{Malkiel}. However, it is possible to generalize this simple deterministic model to incorporate nonlinear positive feedback on the stochastic Black-Scholes model, leading to the concept of stochastic finite-time singularities 
\cite{SornetteAndersen02,Fogedby1,Fogedby2,AndersenSornette04}. Still much work needs to be done on this theoretical aspect. 

Coming back to figure \ref{Fig_HengSeng_Bubbletextbook} , one can also notice that each burst of super-exponential price growth is followed by a crash, here defined for the eight arrowed cases as a correction of more than 15\% in less than three weeks. These examples suggest that the non-sustainable super-exponential price growths announced a ``tipping point'' followed by a price disruption, i.e., a crash.  The Hong-Kong Hang Seng index provides arguably one of the best textbook example of a free market in which bubbles and crashes occur repeatedly: the average exponential growth of the index is punctuated by a succession of bubbles and crashes, which seem to be the norm rather than the exception.

More sophisticated models than (\ref{tgtga}) have been proposed to take into account the interplay between technical trading and herding (positive feedback) versus fundamental valuation investments (negative mean-reverting feedback). Accounting for the presence of inertia between information gathering and analysis on the one hand and investment implementation on the other hand \cite{IdeSornette02}  or between trend followers and value investing \cite{Farmer02}, the resulting price dynamics develop second-order oscillatory terms and boom-bust cycles. Value investing does not necessarily cause prices to track value. Trend following may cause short-term trend in prices, but also cause longer-term oscillations. 

The simplest model generalizing (\ref{tgtga}) and including these ingredients is the so-called log-periodic power law (LPPL) model (Ref.~\cite{Sornettebookcrash03} and references therein). Formally, some of the corresponding formulas can be obtained by considering that the exponent $m$ is a complex number with an imaginary part, where the imaginary part expresses the existence of a preferred scaling ratio $\lambda$ describing how the continuous scale invariance of the power law (\ref{tgtga}) is partially broken into a discrete scale invariance \cite{Sornette1998}. The LPPL structure may also reflect the discrete hierarchical organization of networks of traders, from the individual to trading floors, to branches, to banks, to currency blocks. More generally, it may reveal the ubiquitous hierarchical organization of social networks recently reported \cite{ZhouDunbar05} to be associated with the social brain hypothesis \cite{Dunbar1998}. The simple implementation of the LPPL model 
that we use in section \ref{wthlsjdgb} reads
\begin{equation}
\ln[p(t)] = A + B (t_c -t)^m \left[1 + C \cos \left(\omega \log(t_c-t) +\phi\right)\right]~,~~~{\rm with}~0 < m < 1~, ~{\rm and}~B<0~.
\label{LPPLmodel}
\end{equation}
The constant $A$ is by construction equal to $\ln[p(t_c)]$. The two key parameters are the exponent $m$,
which characterizes the strength of the super-exponential acceleration of the price on the approach to the
critical time $t_c$, and $\omega$, which encodes the discrete hierarchy of accelerated 
``impulse-retracting'' market wave patterns associated with the super-exponential acceleration.
Specifically, the preferred scaling ratio encoding the accelerated oscillations 
is given by $\lambda \equiv e^{2 \pi \over \omega}$ \cite{Sornette1998}.

Examples of calibrations of financial bubbles with one implementation of the LPPL model are the 8 super-exponential regimes discussed above in figure  \ref{Fig_HengSeng_Bubbletextbook}: the 8 small insets at the bottom of figure  \ref{Fig_HengSeng_Bubbletextbook} show the LPPL calibration on the Hang Seng index 
on the bubble phase that preceded each peak. Preliminary tests suggest that the LPPL model provides a good starting point to detect bubbles and forecast their most probable end \cite{Sornettebookcrash03}. Rational expectation models of bubbles a la Blanchard and Watson implementing the LPPL model \cite{JLS1,JLS2,JohansenSornette06} have shown that the end of the bubble is not necessarily accompanied by a crash, but it is indeed the time where a crash is the most probable. But crashes can occur before (with smaller probability) or not at all. That is, a bubble can land smoothly, approximately one-third of the time, according to preliminary investigations \cite{JohansenSornette06}. Therefore, only probabilistic forecasts can be developed. Probability forecasts are indeed valuable and commonly used in daily life, such as in weather forecasts.


\section{A 15 year history of the 2007-???? financial and economic crisis}
\label{wthlsjdgb}

Using the general framework for bubbles and crashes outlined in 
section \ref{thrjgvqpfq}, we now present the evidence on the 
five successive bubbles that developed over the last 15 years. We suggest
that these five bubbles reveal the belief in the ``perpetual money machine''
that characterized this epoch, as discussed in subsection \ref{tgt2ogt4gt}.

Each bubble excess was thought and felt as  ``solved'' by the following excess... 
leading to a succession and combination of mutually reinforcing
unsustainable financial bubbles, preparing the ground for the instabilities
that have been unravelling since 2007. 
The evidence presented in this section is useful to fully appreciate that 
the present crisis and economic recession are to be understood as
the ``hangover'' and consolidation phase following this series of unsustainable excesses.

One should conclude that the extraordinary severity of this crisis is not going to be solved by the same implicit or explicit ``perpetual money machine'' thinking, that still characterize most of the proposed solutions.
``The problems that we have created cannot be solved at the level of thinking that created them.'' said Albert Einstein.

\begin{figure}[H]
\sidecaption
\includegraphics[scale=.45]{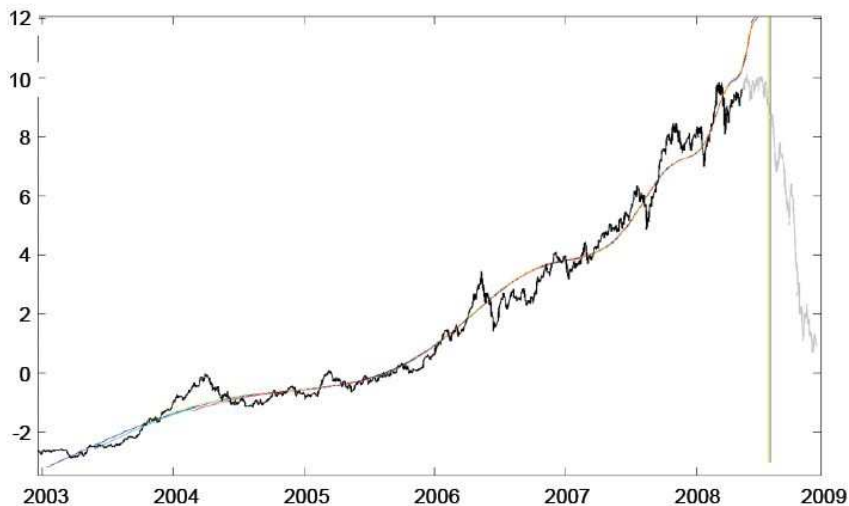}
\caption{First component obtained from a principal component analysis performed on a data set containing, emerging markets equity indices, freight indices, soft commodities, base and precious metals, energy, and currencies.
Source: Peter Cauwels,  Fortis Bank -- Global Markets
}
\label{Fig_global_bubble}        
\end{figure}

We start by presenting the analysis using the LPPL model (\ref{LPPLmodel}) presented in 
subsection \ref{LPPLtreatg} of a global index obtained as follows. 
Starting from time series of emerging market equity indices, freight indices, soft commodities, base and precious metals, energy, and currencies, a principal component analysis (PCA) yields a set of 
principal components that are thought to correspond to common factors acting 
on these time series. The first principal component, which explains the largest
fraction of the covariance of these time series, is shown in figure \ref{Fig_global_bubble},
together with its fit with the LPPL model  (\ref{LPPLmodel}). It is striking to 
observe the overall super-exponential behavior, with a clear change of regime occurring
mid-2008. The following subsections allows us to decompose this overall process
into bubble components.

\subsection{FIRST PHASE: the ITC ``new economy'' bubble (1995-2000)}

\begin{figure}[H]
\sidecaption
\includegraphics[scale=.4]{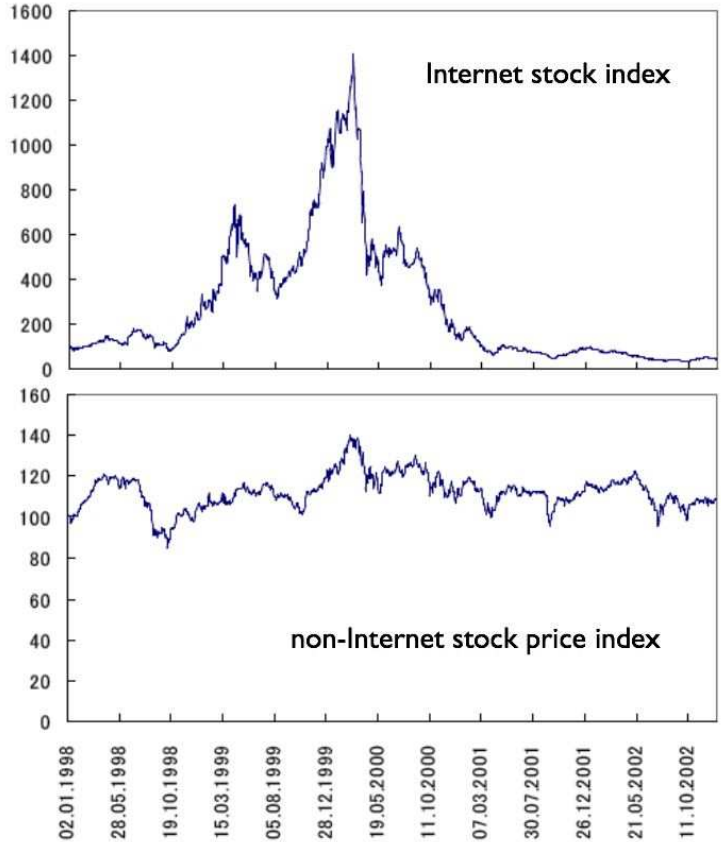}
\caption{
The Internet stock index and non-Internet stock index which are equally weighted as explained in the text. Comparison of the index levels of the Internet index and the non-Internet Stock index for the period 2 Jan. 1998 to 31 Dec. 2002. The two indexes are scaled to be 100 on 2 Jan. 1998. Courtesy of Taisei Kaizoji
}
\label{Fig_Internet_NonInternet}        
\end{figure}

The nature of the ITC bubble is striking when comparing the price trajectories
of two indices constructed on the 500 companies forming the S\&P500 index.
The Internet stock index is an equally weighted portfolio of 100 firms 
related to the Internet. The non-Internet stock price index is an equally weighted 
portfolio made of the remaining 400 ``brick-and-mortar'' companies.
Figure \ref{Fig_Internet_NonInternet} shows that the non-Internet stock price
index remained basically flat from 1998 to 2002, while exhibiting fluctuation
of roughly $\pm 20\%$ over this period. In contrast, the Internet stock index was multiplied
by a factor 14 from 1998 to its peak in the first quarter of 2000, and then shrunk
with a great crash followed by a jumpy decay to below its initial value at the end of 2002.
The contrast between the behavior of these two indices over the same 4 years
interval cannot be more shocking. 

\begin{figure}[H]
\includegraphics[scale=.26]{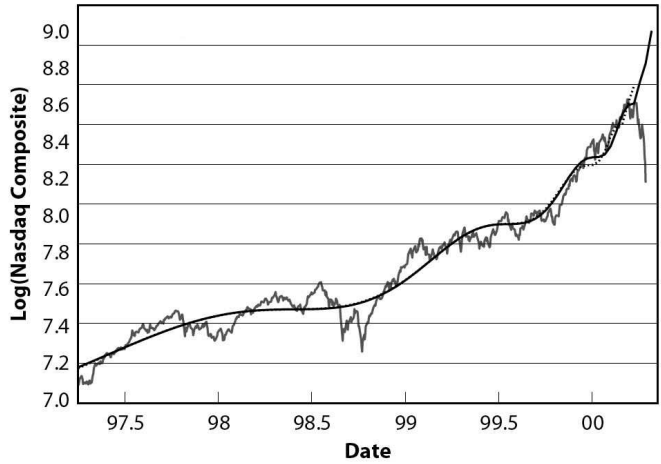}
\caption{Calibration of the LPPL model (\ref{LPPLmodel}) to the Nasdaq 
Composite Indexfrom early 1997 to the end of 1999. 
Reproduced from Ref.~\cite{JohansenSornetteNasdaq00} 
}
\label{Fig_Nasdaq_bubbles}        
\end{figure}

The super-exponential nature of the Nasdaq composite index allows us to 
diagnose this period before 2000 as an unambiguous bubble as first
reported by Johansen and Sornette (2000) \cite{JohansenSornetteNasdaq00}, according 
to the definition presented in subsection \ref{LPPLtreatg}.
Figure \ref{Fig_Nasdaq_bubbles} shows that the logarithm of the
Nasdaq composite index indeed increased with an overall upward
curvature, signaling a super-exponential growth. The calibration of the
LPPL model (\ref{LPPLmodel}) to the Nasdaq index is excellent
(see Ref.~ \cite{JohansenSornetteNasdaq00} for details and statistical tests).

\begin{figure}[H]
\includegraphics[scale=.17]{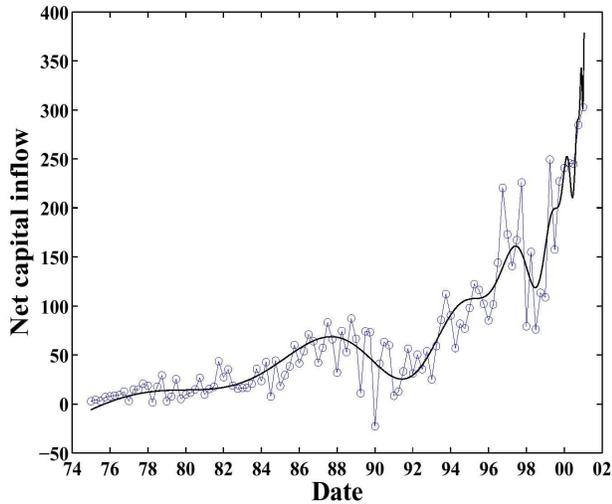}
\caption{Foreign capital inflow in the U.S. during the ITC bubble, illustrating
the growth of the euphoria phase in scenario (\ref{thy;e;ss}) of subsection \ref{hntykgte}.
The smoothed curve shows the fit of the net capital inflow 
by an extension of the LPPL model (\ref{LPPLmodel}) using higher-order
log-periodic components presented in Ref.~\protect\cite{SorZhou04}
}
\label{Fig_Capital-Inflow_Nasdaqbubble}        
\end{figure}

As explained in the scenario (\ref{thy;e;ss}) in subsection \ref{hntykgte},
a typical bubble goes through a period of euphoria. This euphoria
is characterized by an irresistible attraction, in particular, to foreign investors,
who cannot wait to be part of the celebration. This pattern is vividly
observed in the case of the ITC bubble in figure \ref{Fig_Capital-Inflow_Nasdaqbubble},
which shows the flux of foreign capital inflow to the U.S. This inflow almost
reached 400 billion dollars per year at the peak. A significant part of this foreign capital
was invested in the U.S. market to profit from the return opportunities it provided
until 2000. The smoothed curve shows that the net capital inflow can also
be well-fitted by the LPPL model (\ref{LPPLmodel}), yielding 
values for the exponent $m$ and log-frequency $\omega$, which are
consistent with those obtained for other bubbles \cite{Johansen03}.

\subsection{SECOND PHASE: Slaving of the Fed monetary policy to the stock market descent (2000-2003)}
\label{thbjrgwpbgmvq}

\begin{figure}[H]
\includegraphics[scale=.38]{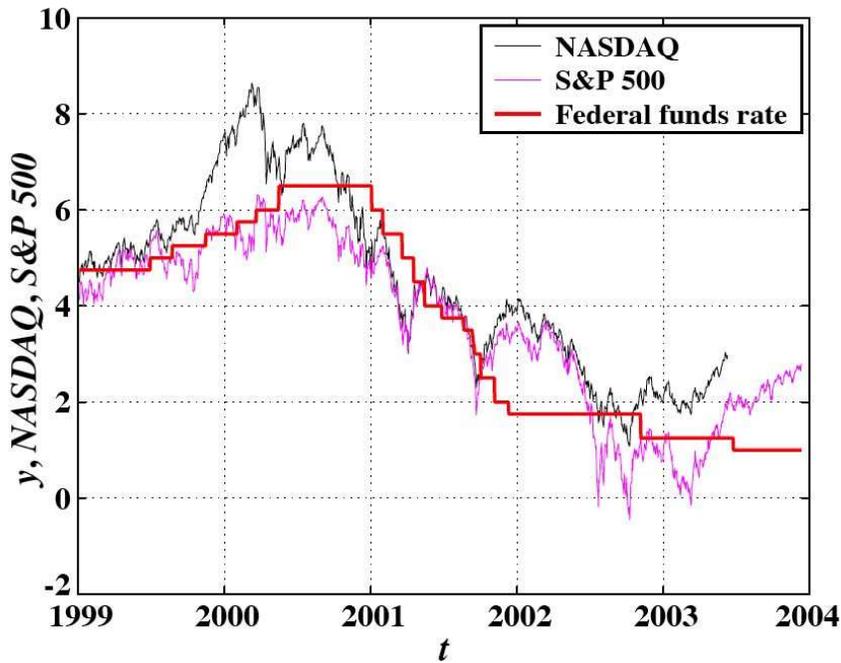}
\caption{
Comparison of the Federal funds rate, the S\&P 500 Index $x(t)$, and the NASDAQ composite $z(t)$, 
from 1999 to mid-2003. To allow an illustrative visual comparison, the indices have been
translated and scaled as follows: $x \to 5x - 34$ and $z \to 10z - 67$.
Reproduced from Zhou and Sornette (2004) \protect\cite{zhousornette2004} 
}
\label{Fig_Slaving_Fed}        
\end{figure}

To fight the recession and the negative economic effects of a collapsing stock market, the Fed engaged in a
pro-active monetary policy (decrease of the Fed rate from 6.5\% in 2000 to 1\% in 2003 and 2004).
Figure \ref{Fig_Slaving_Fed} shows this decrease of the Fed rate and compares
it with the behavior of two U.S. market indices, the S\&P500 and the Nasdaq composite indices.

It is quite apparent that the Fed rate decreased in parallel to the U.S. stock market. 
But did it lead it or lag behind it?  According to common wisdom, the Federal Reserve
control of the leading rate indicator is supposed to influence the stock markets. A decrease
of the Fed rate makes borrowing cheaper, giving more leverage to firms to invest
for the future. As a consequence, this should lead to anticipations of larger future
growth, and hence to larger present market values. Hence, logically, the Fed rate drops
should precede the market losses.

\begin{figure}[H]
\includegraphics[scale=.5]{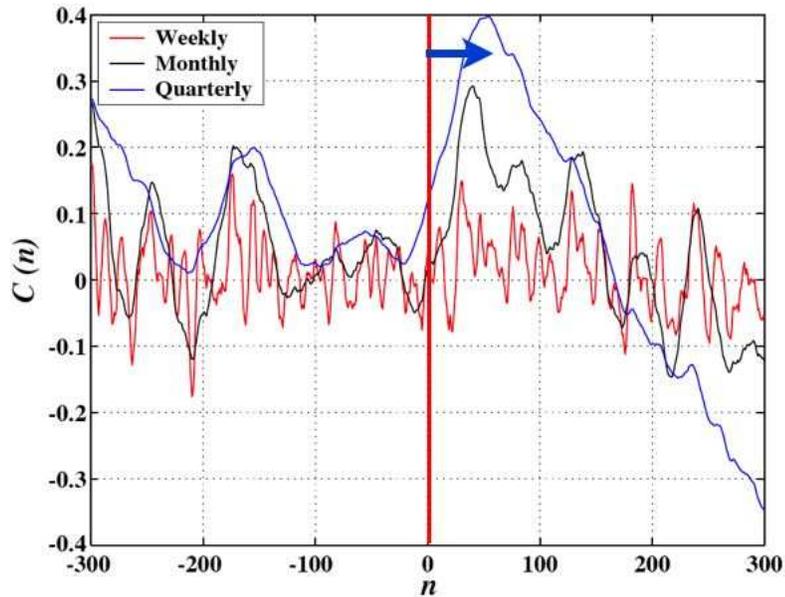}
\caption{``Causal Slaving'' of the U.S. Treasury Bond Yield by the Stock Market Antibubble of August 2000.
The cross-correlation coefficient $C(n)$ between the increments of the logarithm of the
S\&P 500 Index and the increments of the Federal funds rate is shown as a function of time lag $n$ in days. The
three curves corresponds to three different time steps used to calculate the increments: weekly, monthly and
quarterly. A positive lag $n$ corresponds to having the Federal funds rate posterior to the stock market.
The arrow points to this lag.
Reproduced from Zhou and Sornette (2004) \protect\cite{zhousornette2004} 
}
\label{Fig_correl_lag_Fed}        
\end{figure}

We check this prediction by
showing in figure \ref{Fig_correl_lag_Fed} the cross-correlation between 
the returns of the S\&P 500 index and the increments of the Federal funds rate as a function of time lag.
The remarkable result is that the Fed rate decreased with a robustly determined
lag of about $1-2$ months behind the on-going loss of the S\&P500. This reverse 
causality suggests that the Fed monetary policy has been influenced significantly by (or ``slaved'' to) 
the vagaries of the stock market. 

In 2003, Fed Chairman A. Greenspan argued that the Fed needed to set low
interest rates to prevent the U.S. economy from deteriorating so much that 
it would follow a deflationary spiral, often referred to as a ``liquidity trap'' \cite{Krugman_liquidy_trap98}, 
a situation in which conventional monetary policy loses all traction.
Greenspan's critics continue to debate about the
influence of the exceptionally low Fed rates in 2002 and 2003 that are thought to have caused
an extraordinary real estate bubble... that led eventually to the 2007-???? crisis and recession.

\begin{figure}[H]
\includegraphics[scale=.8]{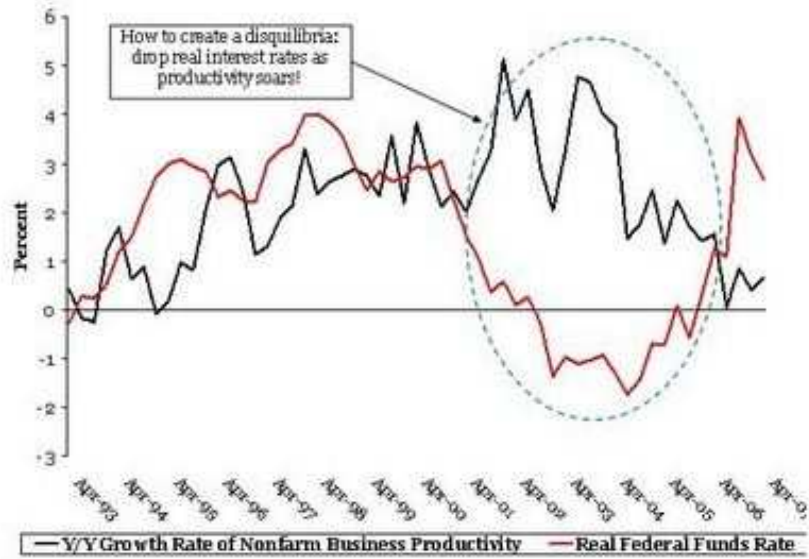}
\caption{Growth rate of the non-farm business productivity compared with the real
Federal fund rate. Reproduced from (Nick Rowe, 2009)
\protect\cite{Rowe-vs-Greenspan}
}
\label{Fig_Greenspan_not_right}        
\end{figure}

Recently, economist Nick Rowe presented a piece of evidence \cite{Rowe-vs-Greenspan}
that is reproduced in figure \ref{Fig_Greenspan_not_right}, which illuminates this debate.
The growth rate of the non-farm business productivity is compared with the real
Federal fund rate. It is apparent that the Fed rate was pushed down at the time
of a surge of productivity gains, not really a deteriorating economy.
This combines with the previous evidence of figure \ref{Fig_correl_lag_Fed}
to support the view that the Federal Reserve has been too obnubilated by
the stock market signals. This is additional evidence that, even in the higher
spheres of finance, the stock market is taking over in shaping economic
and strategic decisions. We mentioned in the introduction of section 
\ref{tktgk3;w} that the impact of financial markets has been growing to basically
dominate strategic decision at the firm level (see also Ref.~\cite{Broekstra} 
for a dramatic example of this trend for the case Royal Ahold firm in the Netherland).
It seems that the monetary authorities are also infected by the same stock market virus. 

\subsection{THIRD PHASE: Real-estate bubbles (2003-2006)  \label{realeatgrgwef}}

The pro-active monetary policy of the Federal Reserve described in the previous subsection,
together with expansive Congressional real-estate initiatives, fueled what can now be rated as one of the most extraordinary real-estate bubbles in history, with excesses on par with those that occurred during the famous real-estate bubble in Japan in the late 1980s. 

\begin{figure}[H]
\includegraphics[scale=.37]{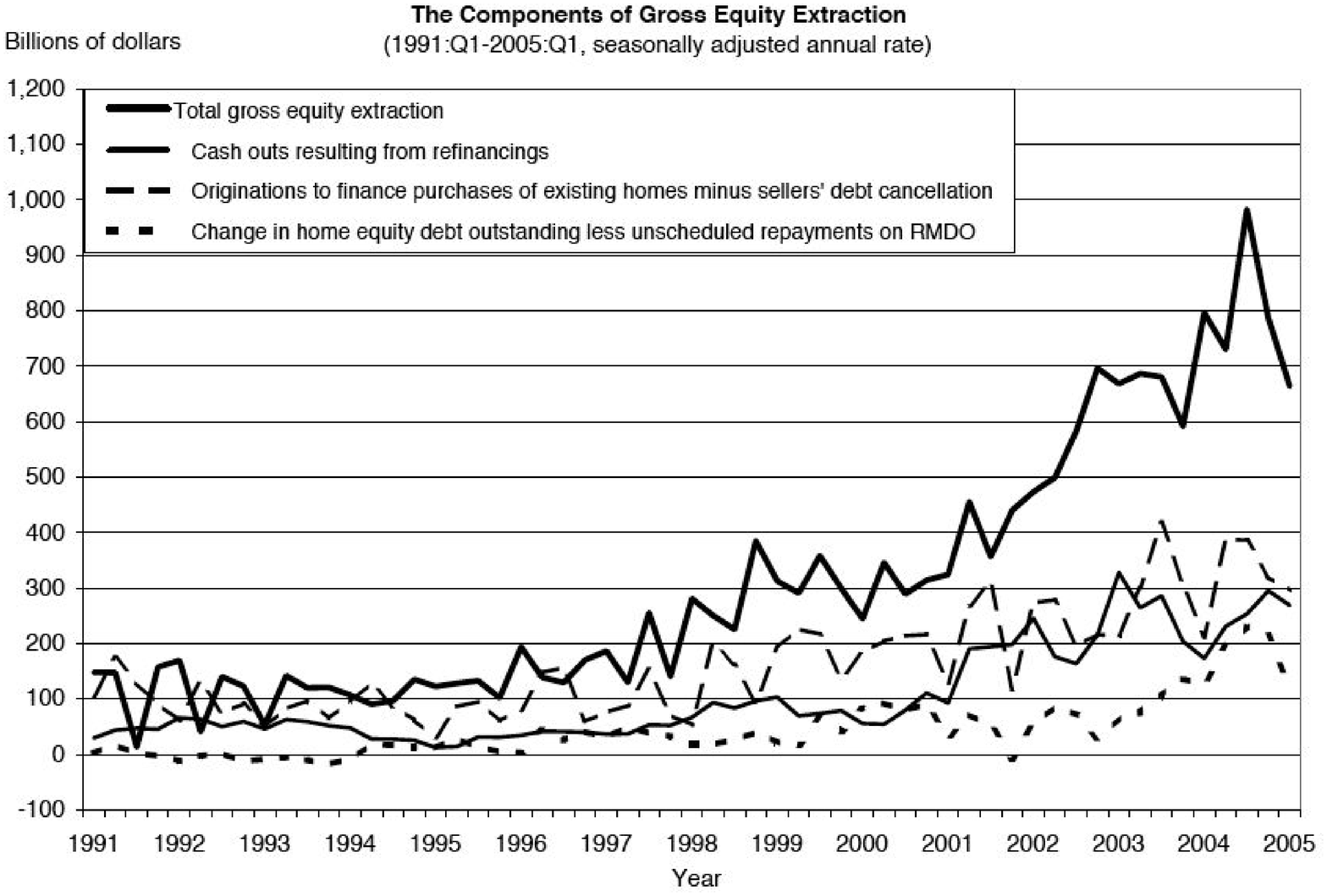}
\caption{Quantification of gross equity extraction by homeowners from their houses,
showing the accelerated growth that spilled over to the economy by fueling
consumption. This figure shows that, over the past decade and a half, 
equity extraction has been closely correlated with realized capital
gains on the sale of homes.
Source: Greenspan and Kennedy (2005) \protect\cite{Greenspan-Kennedy05}
}
\label{Fig_Equity_extraction_Greenspan}        
\end{figure}

\begin{figure}[H]
\includegraphics[scale=.4]{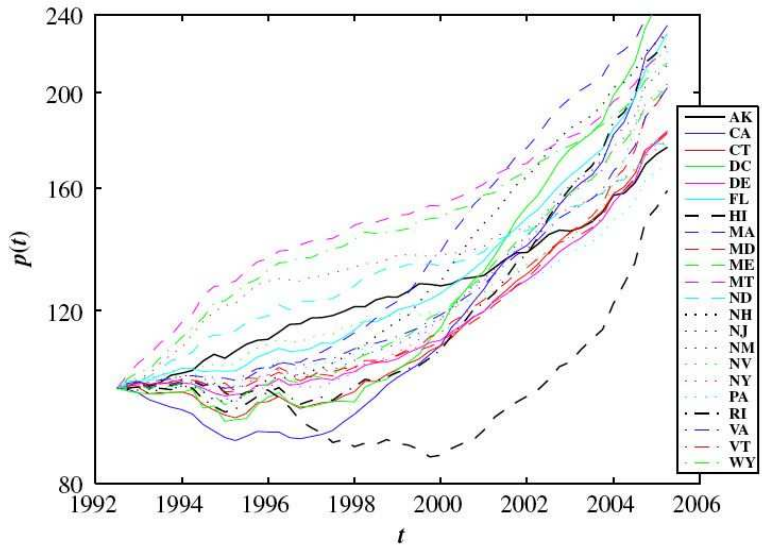}
\caption{(color online) Quarterly average HPI (house price index) in the 21 states and in the district of Columbia (DC)
that were found to exhibit a clear faster-than-exponential growth. For better
comparison, the 22 house price indices have been normalized to 100 at the second
quarter of 1992. The corresponding states symbols are given in the legend. 
Reproduced from Zhou and Sornette (2006) \protect\cite{zhousornette2006}
}
\label{Fig_Real-Estate_USA}        
\end{figure}

As Alan Greenspan himself documented in a scholarly paper researched during his tenure as the Federal Research Chairman at that time \cite{Greenspan-Kennedy05}, the years from 2003 to 2006 witnessed an extraordinary acceleration of the amount of wealth extracted by Americans from their houses as shown in figure \ref{Fig_Equity_extraction_Greenspan}, which parallels the accelerated house price appreciation shown in figure \ref{Fig_Real-Estate_USA}. The negative effects on consumption and income due to the collapse of the first ITC bubble were happily replaced by an enthusiasm and a sense of riches permeating the very structure of US society. 

\begin{figure}[H]
\includegraphics[scale=.4]{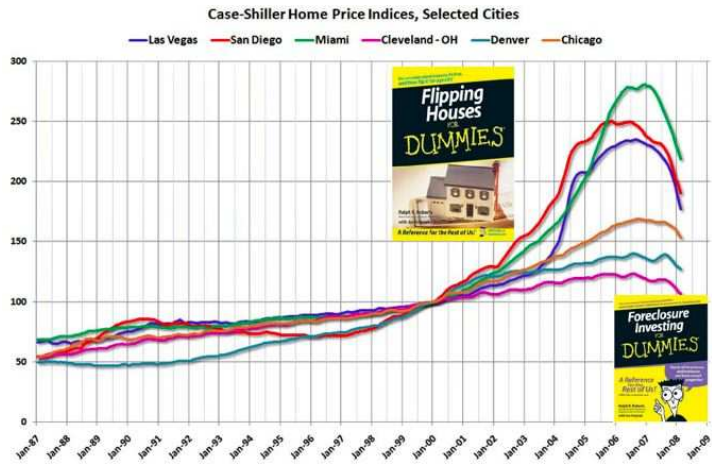}
\caption{Year-over-year price changes for the Case-Shiller composite 10 and 20 indices (through February 2008), and the Case-Shiller and OFHEO National price indices (through Q4 2007).  Adapted from 
\protect\url{http://calculatedrisk.blogspot.com}. The pictures of the two books are put here
to emphasize the dominating sentiment in each phase (see Ref.~\protect\cite{Roehnersor}
for a study of how book sales reflect market bubbles)
}
\label{Fig_realized_Case-Shiller}        
\end{figure}

In June 2005 (proof from the arXiv submission \protect\url{http://arxiv.org/abs/physics/0506027}),
Zhou and Sornette (2006) issued a diagnostic  that about 2/5 of the states of the U.S.
were developing real estate bubbles.
Zhou and Sornette (2006) predicted a peak for most of the US real estate bubbles in mid-2006 
\protect\cite{zhousornette2006}.
The validity of this prediction can be checked in figure \ref{Fig_realized_Case-Shiller}.

It should be noted that the real estate bubble has not been confined to the U.S. but was active in many (but not all)
countries. Exceptions include Germany, Japan, Switzerland and The Netherlands. But the
critical time of the peak of the bubble has been different in different countries. For instance, it
was mid-2004 for the U.K. bubble \cite{zhousornette2003} compared to mid-2006 for the U.S. bubble.

\subsection{FOURTH PHASE: MBS, CDOs bubble (2004-2007)}

\begin{figure}[H]
\includegraphics[scale=.57]{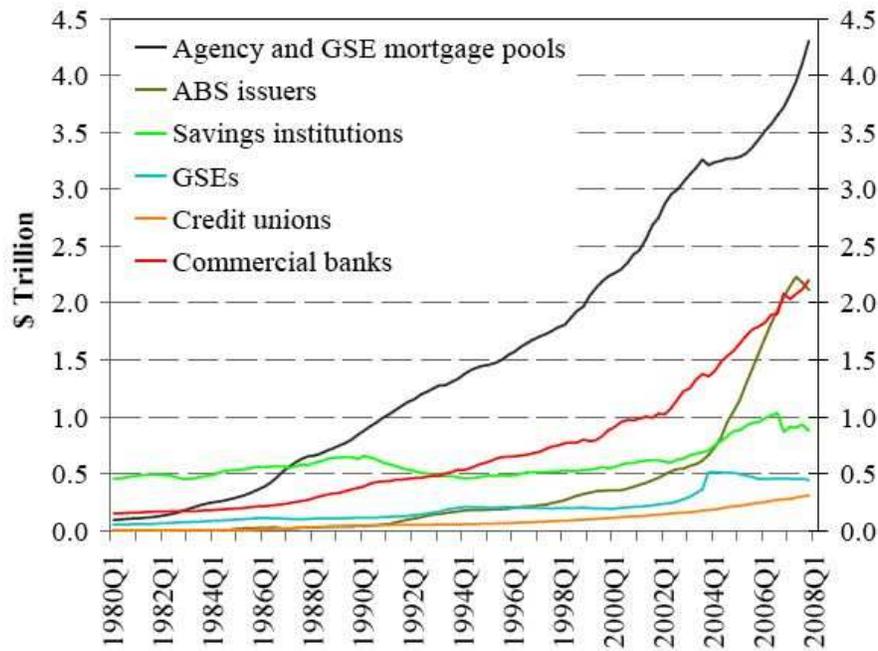}
\caption{Total Holdings of US Home Mortgages
by Type of Financial Institution. Source: Hyun Song Shin, Princeton University
}
\label{Fig_Holdings_mortgages}        
\end{figure}

\begin{figure}[H]
\includegraphics[scale=.47]{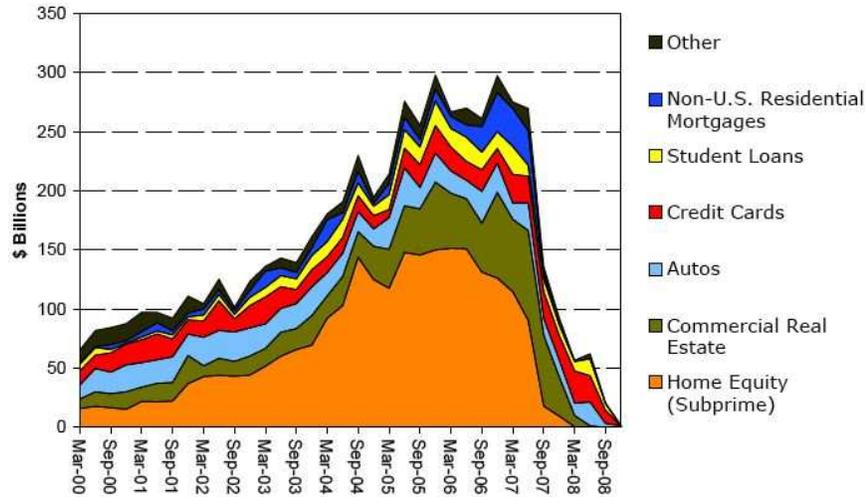}
\caption{New Issuance of Asset Backed Securities
in Previous Three Months. Source:  JP Morgan
}
\label{Fig_ABS-growth}        
\end{figure}

Concomitantly with the real estate bubble, 
both the public and Wall Street were the (sometimes unconscious) actors of a third bubble of subprime mortgage-backed securities (MBS) and complex packages of associated financial derivatives, as
already described in subsection \ref{hjgrld} and now shown in figures \ref{Fig_Holdings_mortgages}  
and \ref{Fig_ABS-growth}.

The growth of the MBS derivatives is exemplified in figures \ref{Fig_Holdings_mortgages}  
and \ref{Fig_ABS-growth}, which respectively show (i) the total holding of mortgage related
securities of different financial institutions and (ii) the accelerated rate of new issuance
of ABS until the peak in March 2007, when the first signs of accelerating loan payment
defaults started to be felt on the MBS.

These two figures \ref{Fig_Holdings_mortgages}  
and \ref{Fig_ABS-growth} clearly illustrate the MBS bubble and its bursting.
In addition, as pointed out by many astute observers, 
many of the MBS were ``fragile'' as they were linked to two key unstable processes: the value of houses and the loan rates. The ``castles in the air'' of bubbling house prices promoted a veritable eruption of investments in MBS, these investments themselves pushing the demand for and therefore the prices of houses -- until the non-sustainability of these mutually as well as self-reinforcing processes became apparent.  

But to be clear; these financial instruments were great innovations which, in normal times, would indeed have provided a win-win situation: more people have access to loans, which become cheaper because banks can sell their risks to the supposed bottomless reservoirs of investors worldwide with varying appetites for different risk-adjusted returns. 

The problem is that the MBS and collateral debt obligations (CDO) constituted new types of derivatives.
Their complexity together with the lack of historical experience may have provided the seed 
for unrealistic expectations of low risks and large returns. Actually, this is part of a larger debate
on the role of financial derivatives. 

Many financial economists hold that derivatives serve
a key role of making markets more complete, in the sense that more states of the world
can be hedged by a corresponding asset. As a consequence, financial markets become
more efficient and stable. Perhaps the most influential proponent of this view has been
Alan Greenspan himself. For more than a decade, Greenspan has fiercely objected whenever derivatives have come under scrutiny in Congress or on Wall Street. ``What we have found over the years in the marketplace is that derivatives have been an extraordinarily useful vehicle to transfer risk from those who shouldn't be taking it to those who are willing to and are capable of doing so,'' Mr. Greenspan told the Senate Banking Committee in 2003. ``We think it would be a mistake'' to more deeply regulate the contracts, he added.
``Not only have individual financial institutions become less vulnerable to shocks from underlying risk factors, but also the financial system as a whole has become more resilient.'' -- Alan Greenspan in 2004.

Others disagree. The well-known financier G. Soros avoids using derivatives ``because we donÕt really understand how they work.'' Felix G. Rohatyn, the investment banker whose action was instrumental during New York financial turmoils in the 1970s, described derivatives as potential ``hydrogen bombs.''
And,  in the 2002  Berkshire Hathaway annual report, Warren E. Buffett observed that derivatives were ``financial weapons of mass destruction, carrying dangers that, while now latent, are potentially lethal.''

These statements have recently been given theoretical support in new out-of-equilibrium
models of financial markets in which it is found that, paradoxically, on the one hand the proliferation of financial 
instruments tends to make the market more complete and efficient by 
providing more means for risk diversification, while at the same time this proliferation of financial instruments erodes systemic stability as it drives the market to a critical state characterized by large susceptibility, strong fluctuations and, enhanced correlations among risks \cite{Brocketal08,Marsili08,Marsilietal08}

\subsection{FIFTH PHASE: Stock market bubble (2004-2007)}

\begin{figure}[H]
\includegraphics[scale=.4]{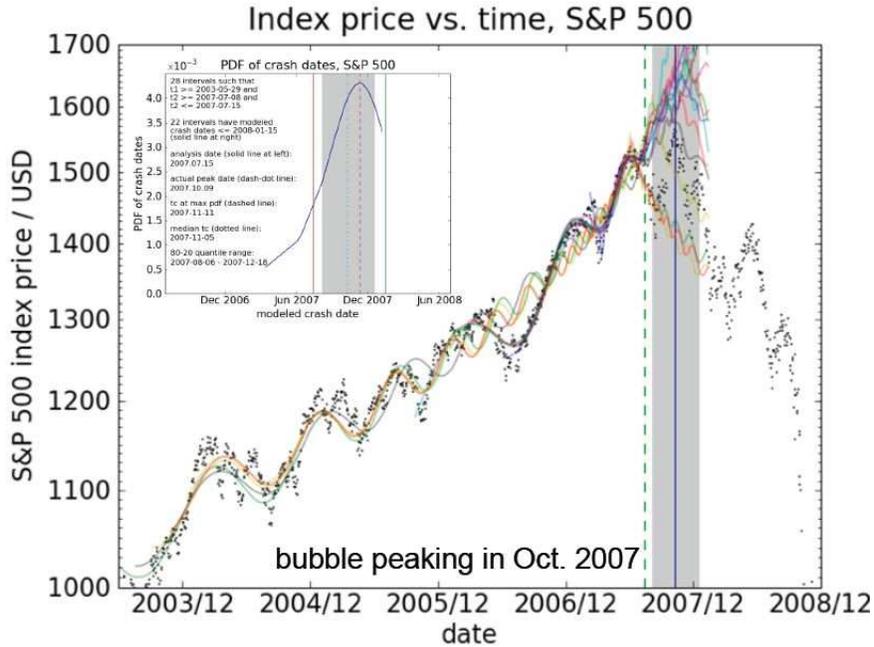}
\caption{S\&P500 index (in logarithmic scale) shown as dots as a function of time.
The dashed vertical line shows the last observed time $t_{\rm last}$ used to perform the calibration of the 
LPPL model (\protect\ref{LPPLmodel}) and the different smoothed curves
correspond to different estimations obtained with distinct time windows
extending no later that $t_{\rm last}$. The grey zone corresponds to the $80\%$
confidence interval for the predicted critical time $t_c$ of the end of the bubble.
The inset shows the probability density function of the predicted $t_c$'s
}
\label{Fig_SP500_2007bubble}        
\end{figure}

The exuberant real-estate market and MBS bubbles spilled over to the stock market.
Figure \ref{Fig_SP500_2007bubble}  shows the S\&P500 index (in logarithmic scale)
as a function of time. A clear upward overall upward curvature
can be observed, which is characteristic of a super-exponential growth.
The LPPL calibration confirms the existence of bubble characteristics. 

\subsection{SIXTH PHASE: Commodities and Oil bubbles (2006-2008)}

\begin{figure}[H]
\includegraphics[scale=.4]{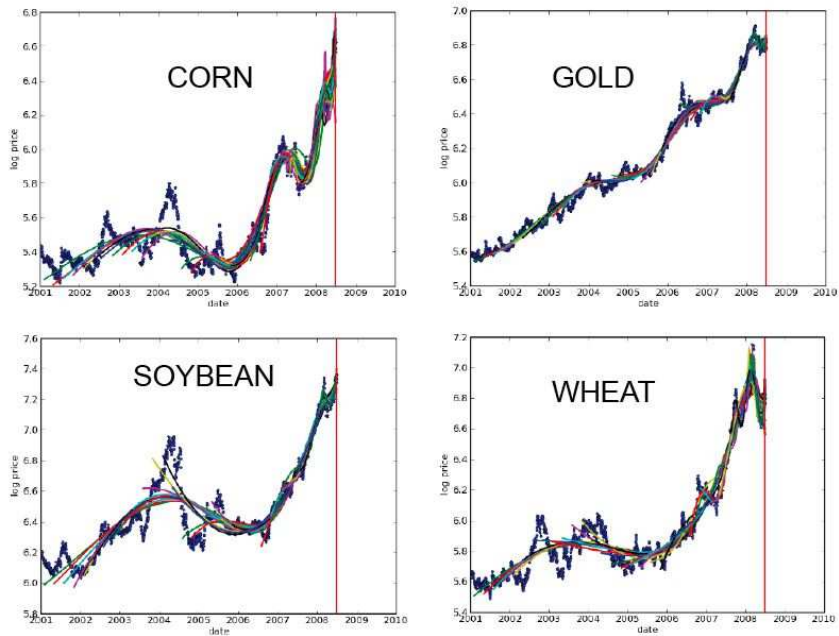}
\caption{Calibration of the LPPL model (\protect\ref{LPPLmodel}) to 
the time series of four commodities (corn, gold, soybean and wheat)
expressed in U.S. dollars
}
\label{Fig_commodity_bubbles}        
\end{figure}

\begin{figure}[H]
\includegraphics[scale=.4]{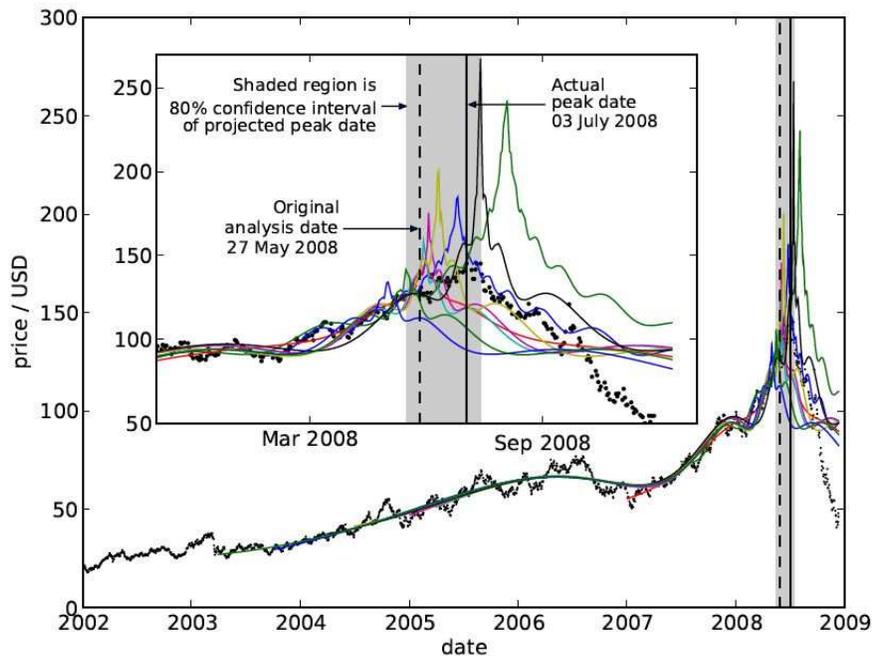}
\caption{Typical result of the calibration of the simple LPPL model to the oil price in US\$ in shrinking windows with starting dates $t_{\rm start}$ moving up towards the common last date $t_{\rm last} =$ May 27, 2008.
Reproduced from Sornette et al. (2009) \protect\cite{SornetteWoodZhou09}
}
\label{Fig_Oil_bubble_2008}        
\end{figure}

As explained in subsection \ref{realeatgrgwef} and shown in figure \ref{Fig_Equity_extraction_Greenspan},
the growth of the real-estate bubble, of the MBS bubble, and of the stock market bubble 
led to a huge extraction of wealth or, in other words, the creation of a lot of wealth and of money.
Both money creation and wealth increase led to higher demand in all things that 
can be consumed. In fact, the
demand has been accelerating on basic commodities,  which developed clear bubble characteristics,
as shown in figures \ref{Fig_commodity_bubbles}  and \ref{Fig_Oil_bubble_2008}.

Oil prices exhibited a record rise, whose starting phase can be traced to 2003 according to our analysis \cite{SornetteWoodZhou09}, followed by a spectacular crash in 2008. The peak of \$145.29 per barrel was set on July  3, 2008 and a recent low of \$40.81 was scraped on December 5, 2008, a level not seen since 2004. On May 27, 2008, we addressed the question of whether oil prices were exhibiting a bubble-like dynamics, which may be symptomatic of speculative behavior, using our techniques based on statistical physics and complexity theory \cite{SornetteWoodZhou09}. Thorough analysis of our May 27, 2008 experiment predicted a peak within a 80\% confidence interval between May, 17 2008 and July 14, 2008. The actual observed `crash', where prices began a long downward trend, began on the last day of this period.

\section{Thoughts on resolution of the crisis and its aftermath}

\subsection{Summary}

We have presented evidence that the fundamental cause of the 
unfolding financial and economic crisis lies in the accumulation of at least $5$
bubbles whose interplay and mutual reinforcement has led to an illusion
of the ``perpetual money machine'' allowing financial institutions to extract
wealth from an unsustainable artificial process.

The path from MBS problem to a global World recession and to potentially
even higher risks can be outlined as follows:
drop in confidence $\to$ drop in demand $\to$ major recession $\to$ interaction between a deep recession and a
weakened financial system $\to$  increased risk of trade wars
$\to$ collapse of global commodity prices and, thus, revenues for
low-income countries $\to$ global instability.

In March 2007, the first visible signs of a worrisome increase in default rates appeared,
followed in the summer of 2007 by the startling realization by the financial community
that the problem would be much more serious. In parallel with an acceleration of the
default of homeowners and a deteriorating economic outlook, the second part of 2007 and
first part of 2008 saw a rapid increase, punctuated by dramatic bankruptcies, in the
estimated cumulative losses facing banks and other investment institutions; from initial
guesses in the range of tens of billions, to hundreds of billions, then to more than a
trillion of U.S. dollars. The U.S. Federal Reserve and U.S. Treasury stepped up their actions in proportion to the
ever-increasing severity of the uncovered failures, unaware of the enormity of the
underlying imbalances, or unwilling to take the measures that would address the full
extent of the problem, which only now (one hopes) has revealed its enormous systemic
proportions. Let us be blunt: Government has been blissfully unaware of the 
predictable effect of the piling up these bubbles, as each one appeared to displace the
problems induced by the previous one. Monetary easing, the injection of liquidity,
successive bailouts -- all address symptoms and ignore the sickness.

The sickness is the cumulative excess liability present in all the sectors of the US
economy: debts of U.S. households as a percentage of disposable income at around
130\%, those of U.S. banks as a percentage of GDP currently around 110\%, U.S.
Government debt at 65\% of GDP, corporate debts at 90\% GDP, state and local
government debts at 20\% GDP, unfunded liabilities of Medicare, Medicaid, and Social
Security in the range of 3-4 times GDP. Such levels of liabilities in the presence of the
bubbles have produced a highly reactive unstable situation in which sound economic
valuation becomes unreliable, further destabilizing the system. This sickness has only
been worsened by measures that disguise or deny it, like the misapplied innovations of
the financial sector, with their flawed incentive structures, and the de facto support of an
all-too willing population, eager to believe that wealth extraction could be a permanent
phenomenon. On the sustainable long term, the growth of wealth has to be equal to the
actual productivity growth, which is about 2-3\% in real value on the long term in
developed countries.

Acting on the ``theory'' that crises can be stopped if confidence is restored by stopping
the hemorrhage of MBS losses, and on the basis of the diagnostic that the 
financial crisis was a ``liquidity'' problem,
central banks and governments have actively
intervened to combat the observed swing in risk taking shown by banks. One
set of measures amount basically to attempt stopping the devaluation of the MBS
assets held by financial institutions. Another set of measures tries
to remove the so-called ``toxic'' assets and park them in vehicles
mainly owned by the government and the Federal Reserve, until 
their values rebound, one hopes. Most measures attempt to encourage
more consumption and spending.

The evidence discussed above suggests strongly 
 that this approach constitutes a fundamental error because it misses the crucial point
about the cause of the ``losses.'' The losses are not just the downturn phase of a
business cycle or the liquidity response to an unlucky exogenous shock.
They express a simple truth that is too painful to accept for most, that
previous gains were not real, but just artificially inflated values that have bubbled in
the financial sphere, without anchor and justification in the real economy. In the last
decade, banks, insurance companies, Wall Street as well as Main Street -- we all have
lured ourselves into believing that we were richer. But this wealth was just the result
of a series of self-fulfilling bubbles: in the USA and in Europe, we had the Internet
bubble (1996-2000), the real-estate bubble (2002-2006), the MBS bubble (2002-
2007), an equity bubble (2003-3007), and a commodity bubble (2004-2008), each
bubble alleviating the pain of the previous bubble or supporting and justifying the
next bubble. In a word, the overgrowth of the ``financial economy''
compared with that of the real economy is reminiscent of the fabled frog
seized by a jealous desire to equal the ox in size \cite{frog-ox}.

The painful consequence of this brutal truth is that trying to support the level of valuation based on these bubbles is tantamount to continue supporting the ``perpetual money machine'', which was the cause of the problem. Worse, it misuses scarce taxpayer resources, increasing long-term debts and liabilities, which are already at dangerous levels in many countries. 

Encouraging over-spending to solve a crisis due to over-spending is an ill-advised approach.

There are no silver bullets, but the following concepts should constitute part of a basis for a pragmatic approach to the crisis.

\subsection{Trust! Why it Has Been Lost and How to Regain It}

The on-going credit crisis and panic shows that financial price and economic value
are based fundamentally on trust; not on fancy mathematical formulas, not on subtle
self-consistent efficient economic equilibrium; but on trust in the future, trust in
economic growth, trust in the ability of debtors to face their liabilities, trust in
financial institutions to play their role as multipliers of economic growth, trust that
your money in a bank account can be redeemed at any time you choose. Usually, we
take these facts for granted, as an athlete takes for granted that her heart will continue
to pump blood and oxygen to her muscles. But what if she suffers a sudden heart
attack? What if normal people happen to doubt banks? Then, the implicit processes of
a working economy -- all we take for granted -- starts to dysfunction and spirals into a
global collapse.

Because of the failed governance, the crisis has accelerated, now in a burst of such
intensity that it has forced coordinated actions among all major governments. While
their present involvement may restore short-term confidence, much more is needed to
address the depth of the problem. At one level, the loss of trust between financial
institutions stems from the asymmetric information on their suddenly escalading
counter-party risks, making even the most solid bank be perceived as a potential
candidate for default, leading to credit paralysis. This can be addressed by financial
measures and market forces. At a deeper level, people in the street have lost
confidence, by observing a succession of problems, timidly addressed by decision
makers proposing ad hoc solutions to extinguish the fire of the day (Bear Stearns,
Fannie, Freddie, AIG, Washington Mutual, Wachovia), with no significant result and
only more deterioration.

Nothing can resist loss of trust, since trust is the very foundation of society and
economy. That people haven't yet made a run on the banks is not, given today's
insurance policies against the catastrophes of the past, sufficient indication to the
contrary. In fact, there has been already invisible run on the banks, as electronic
and wire transfers have been accelerating in favor of government-backed Treasury
bills. A significant additional impediment to the restoration of public trust is that the
Fed, Treasury and concerted government actions are perceived as supporting
the notion that ``gains are private while losses are socialized.''

Present actions attempt to stabilize the financial sector by making governments,
therefore taxpayers, the lenders and buyers of last resort. But collectively people are
more intelligent than governments and decision makers think. They know that
governments, in particular in the West, have not saved (counter-cyclically a la
Keynes) during the good years\footnote{In this respect, note the information from Reuters, Santiago,
March 27, 2009, reporting that Chile's President Michelle Bachelet unwittingly embarrassed British Prime Minister Gordon Brown when she said Chile had put aside money during good economic times to help it through the downturn. ``I would say that because of our decision during ... the good times in copper prices, we decided to save some of the money for the bad times and I would say that policy today is producing good results.''},
and they thus wisely doubt their prudence during the
bad ones. They suspect that their governments will eventually extract the needed
capital from them. They suspect intuitively that the massive measures taken to support
the financial world will do little to help general economies of the US, Europe and the
rest of the world.

How to restore trust?  This is a long, tedious and fragile process.
We suggest that governing bodies must for once play to the intelligence of the crowd. What
needs to be done is to explain truthfully (is this possible?) the true cause of
the problems: the more than one decade of excesses and the successive and
inter-related bubbles, the fact that the liabilities are enormous and that the budget has
in the end to be balanced, that accelerating borrowing on the future cannot be a
sustainable strategy. As humans, we are more inspired to trust when failures are
acknowledged than when blame is shifted to someone else. This is the core reason
why going to the fundamental source of the problems may be part of the solution in
restoring confidence in the existence of a solution at minimal cost.

Second, the issue of fairness is essential for restoring confidence and support. 
There is an absolute need to
rebuild that confidence, and this applies also to the regulators. This requires new
strong regulations to deal with conflicts of interest, moral hazard, and to enforce the
basic idea of well-functioning markets in which investors who took risks to earn
profits must also bear the losses. For instance, to fight the rampant moral hazard that
fueled the bubbles, share-holders should be given ``clawback'' permission, that is, the
legal right to recover senior executive bonus and incentive pay, that proved to be ill-founded.
In addition, many of the advisors and actors of the present drama have
vested interest and strong conflict of interests. This particularly
the case in the U.S, for the Fed, the Treasury and the
major banks acting on behalf or with the approval of the Treasury. An independent
(elected?) body would be one way to address this problem, ensuring separation of
interest and power.

\subsection{Short-term: melting the cash flow freeze}

The most immediate issue is to address the cash flow freeze imposed by banks,
with their newfound overly restrictive lending rules, on companies and households.
This cash flow problem bears the seed of a spiraling recession of catastrophic
amplitude, which has no fundamental reason to develop, except as an unwanted
consequence of pro-cyclical feedbacks aggravating a necessary correction that should
only be confined to the financial sphere. Here, the central banks and governments
should show creativity in ensuring that small and medium size companies have access
to monthly liquidity, to allow them to continue producing and hiring. This is the issue
that has been by far the most under-estimated and which requires a fast and vigorous
solution. 

In addition to providing lending facilities to banks conditional on serving
their natural multiplier role in the economy, special governmental structures could be
created with a finite lifetime, with the mandate to provide liquidity to the real
economy, bypassing the reluctant banks. Note that this procedure should not
necessarily be used to bailout some large badly managed companies in some industry
sectors, when in obvious need of restructuring. Crises are often opportunities for
restructuring, which provide increased benefits in the future as some cost in the
present.

B. Lietaer has proposed a different approach based on the use of a 
complementary currency to help make the network of financial interactions
between companies
more robust \cite{Lietaeretal}, in analogy with ecological networks which derive their
resilience from the multiple network levels they are built on.
As a first candidate for this complementary currency could be a professionally run
business-to-business system of payments on the model of the WIR system\footnote{The Swiss Economic Circle (Wirtschaftsring-Genossenschaft or WIR) is an independent complementary currency system in Switzerland that serves small and medium-sized businesses. It was founded in 
1934 by businessmen Werner Zimmermann and Paul Enz as a result of currency shortages after the stock market crash of 1929 and the great recession. ``Its purpose is to encourage participating members to put their buying power at each other's disposal and keep it circulating within their ranks, thereby providing members with additional sales volume.'' Cited from Wikipedia}, which has
been successfully operational for 75 years in Switzerland, involving a quarter of all the businesses
in that country. This system has been credited by J. Stodder as a
significant counter-cyclical stabilizing factor that 
may contribute to the proverbial stability of the Swiss economy \cite{Stodder}.

\subsection{Long-term: growth based on returning to fundamentals and novel opportunities}

Long-term economic stimulation programs are needed on a large scale,
probably a few percent of GDP, with pragmatic adaptive tuning as the crisis unfolds.
They should focus on the fundamentals of wealth growth: infrastructure, education
and entrepreneurship, with the goal of promoting productivity growth and the creation
of new real economic sources of wealth. Many studies demonstrate for instance a
direct impact of machinery equipment on economic growth. 

Similarly, by many
metrics, the quality of education in the USA and to a lesser degree in Europe has been
degrading in the recent decades. This crisis is an opportunity to go back to the
fundamentals of the roots of long-term sustainable wealth creation. These stimulation
programs offer an opportunity to adapt and develop new infrastructure which are
more energy and pollution efficient, thus promoting the development of new industry
sectors such as wind energy, electricity storage, nuclear waste processing and
recycling and so on. 

Given growing evidence that mankind is facing global challenges
for its sustainability on the finite Earth, the financial-rooted crisis offers a chance for
using its associated political capital to make bold choices to steer an environmentally
friendly economic development. Governments are best in their role of risk takers
investing in long-term R\&D projects that provide the support for innovations that
industry can build upon to provide increased prosperity in the future.

\subsection{The financial sphere, bubbles and inflation}

One has to accept the need for an abrupt deflation of the financial sphere. And
for the future, mechanisms should be designed to control the over-growth of the
financial economy to ensure better stability. When functioning properly, the financial
world provides many services such as efficient access to funding for firms,
governments and private people. Furthermore, it works as an effective storage of
value, which should reflect the ``real economy.'' But the extraordinary growth of the
component of wealth associated with the financial world has been artificial and
based on multipliers amplifying the virtual fragile components of wealth. 
Objective measures and indicators can be developed to
quantify the ratio of wealth resulting from finance compared with the total economy.
For instance, when it is assessed that, on average, about 40\% of the income of major US
firms result from financial investments, this is clearly a sign that the US economy is
``building castles in the air.'' In the academic literature, this is related to 
the concept of ``financialization''\footnote{see
\url{http://en.wikipedia.org/wiki/Financialization}}, according to which
profit making occurs increasingly through financial channels, and shareholder value
tends to dominate corporate governance.

The way we think of inflation also needs to be re-evaluated. For instance, a house
price appreciation does not just mean that you are more wealthy as a homeowner; it
also implies that you need more dollars or euros to buy one unit of habitation
compared to units of food, vacation or university tuition. From this vantage, it is part
of inflation. While already considered in present measure of the so-called
consumer price index (CPI), its weight and impact need to be re-evaluated.
We propose that real-estate and equity indices should be incorporated as
constituents of inflation metrics, of course with adequate consideration for the
hedonic gains\footnote{Governments use so-called hedonic regression in computing their CPI
to take quality changes into account.}. A financial ratio index could be created to follow
a broader definition of inflation useful for central bank monitoring, which includes
the growth of total fixed assets, working capital,  excess supply of money and so on.

With such tools, monetary policy with inflation targets will provide natural
partial control over some of the asset bubbles at the origin of the present financial
crisis. Guidelines could be drawn to flag warning signals to central banks and
governments when the ratio of the financial wealth compared with the real economy
value grows above a bracket that could be defined from a consensus among
economists and actions could be taken to moderate the growth of this ratio. These
indicators should be the key targets of modern central banks.

Central banks and governments should step in to support financial institutions,
but only under fair conditions that ensure that stockholders and lower priority debt
holders support the consequences of the losses, avoiding the privatization of gains and
socialization of losses. Different technical mechanisms have been proposed by
financial economists, which serve this goal, safeguarding the interest of the taxpayers
on the long term.

As a final point on the issue of the size of the financial sphere, the 
first author is a happy professor
teaching financial economics to a growing corpus of students in a World-renowned
technical university. We are however worried by the growing flood of civil, mechanical,
electrical and other engineers choosing to become transfuges and work in finance:
Is this another bubble in the making? Finance will not solve the many problems mentioned
above. Creativity and entrepreneurship occurring in the real economy and the real
world need to be better rewarded.

\subsection{Recipes for a more robust and sustainable World}

The present crisis is illustrating the accelerating fragility of society. We believe that this is just a foreshock of much more serious jolts to come on times scales of just one or two decades. In this respect, we refer to 
Johansen and Sornette (2001) \cite{JohSor01} and to to Chapter 10 of 
Sornette (2003) \cite{Sornettebookcrash03}, in which the analysis 
of Earth's human population, its economic output and global stock market capitalization
suggest a transition to a completely new regime circa 2050, over a time
scale of several decades around that date.

However, now is an opportunity to build a more resilient World. Recipes are known. They involve the need for variety, redundancy, compartments, sparseness of networks and consequences of the unavoidable delays and synchronization of actions. This ``robustness'' approach is well exemplified by a conservative investment approach based on (i) understanding the vehicles and firms in which one invests (which contrasts with the opaqueness of the MBS investments) and (ii) keeping capital
even under extraordinarily adverse conditions. This strongly contrasts with standard financial practices based on estimated likelihoods for meeting obligations and short-term gains. This requires fundamentally new design in infrastructures and in regulations. The task is complex, but realizing and formulating it is a major step that should be followed by a vigorous program at the international level, based on multidisciplinary task forces that are well-funded and empowered with authority. Leading countries should start at their domestic level to nucleate the process internationally.

Beyond the immediate concerns, we need to keep in mind the big picture, that this time is a unique opportunity for change and improvement. The present crisis should be exploited to start developing a genuine culture of risks, which should be obligatory training for managers in governments, in regularity bodies, and in financial institutions. One little discussed reason for the present crisis was indeed the lack of adequate education of top managers on risks in all its dimensions and implications. This is also the time that a culture of risk should start permeating the public at large. In the 21st century, ``linear'' and ``equilibrium'' thinking should be replaced by a growing appreciation of the inter-connectivity and feedbacks of the complex systems we deal with, which creates shocks -- with opportunities.

\end{document}